\documentclass[12pt]{article}
\usepackage{epsfig}
\newcommand{\be}{\begin{equation}}
\newcommand{\ee}{\end{equation}}
\newcommand{\bea}{\begin{eqnarray}}
\newcommand{\eea}{\end{eqnarray}}
\newcommand{\Fh}[2]{\,{}_#1F_#2}
\newcommand{\Fs}[3]{\!\!\left[\begin{array}{c}#1\,;\\#2\,;\end{array}#3\right]}

\newcommand{\Ffp}[2]{\Fs{#1}{#2}{\frac{q^2}{4m^2}}}
\newcommand{\FfP}[2]{\Fs{#1}{#2}{\frac{p^2}{4m^2}}}

\newcommand{\Ffx}[2]{\Fs{#1}{#2}{x}}
\newcommand{\Ffz}[2]{\Fs{#1}{#2}{z}}
\newcommand{\Ffw}[2]{\Fs{#1}{#2}{w}}
\newcommand{\Ffe}[2]{\Fs{#1}{#2}{\frac{z}{z-1}}}

\newcommand{\Fzz}[2]{\Fs{#1}{#2}{1-z}}
\newcommand{\Fiz}[2]{\Fs{#1}{#2}{\frac{1}{z}}}
\newcommand{\Foz}[2]{\Fs{#1}{#2}{1-\frac{1}{z}}}

\newcommand{\Fde}[2]{\Fs{#1}{#2}{\frac{m_i^2}{r_{ij}}}}
\newcommand{\Fdz}[2]{\Fs{#1}{#2}{\frac{m_j^2}{r_{ij}}}}

\newcommand{\Fdee}[2]{\Fs{#1}{#2}{1-\frac{m_i^2}{r_{ij}}}}
\newcommand{\Fdzz}[2]{\Fs{#1}{#2}{1-\frac{m_j^2}{r_{ij}}}}

\newcommand{\Frmi}[2]{\Fs{#1}{#2}{\frac{r_{ij}}{m_i^2}}}
\newcommand{\Frmj}[2]{\Fs{#1}{#2}{\frac{r_{ij}}{m_j^2}}}

\newcommand{\Fwz}[2]{\Fs{#1}{#2}{\frac{w-z}{1-z}}}
\newcommand{\Fwzz}[2]{\Fs{#1}{#2}{\frac{t(y-z)}{1-tz}}}

\newcommand{\Fpm}[2]{\Fs{#1}{#2}{1-\frac{p^2}{4m^2}}}
\newcommand{\Fdi}[2]{\Fs{#1}{#2}{1-\frac{r_{ij}}{m_i^2} }}
\newcommand{\Fdj}[2]{\Fs{#1}{#2}{1-\frac{r_{ij}}{m_j^2} }}

\newcommand{\Fdf}[2]{\Fs{#1}{#2}{\frac{r_{ij}}{r_{ijk} }}}
\newcommand{\Fdg}[2]{\Fs{#1}{#2}{1-\frac{p_{12}^2}{4 m^2} }}
\newcommand{\Fbf}[2]{\Fs{#1}{#2}{\frac{r_{ijk}}{r_{ijkl} }}}
\newcommand{\Fxy}[2]{\Fs{#1}{#2}{\frac{x}{y}}}

\begin{document}

%=========== title page ======================
\thispagestyle{empty}
\onecolumn
\date{\today}
\vspace{-1.4cm}

\begin{flushright}
DESY-03-033\\
BI-TP-2003/05 \\
SFB/CPP-6
\end{flushright}

\vspace{0.5cm}

\begin{center}

{\LARGE {\bf
A new hypergeometric representation of one-loop scalar  integrals
in $d$ dimensions
}\vglue 8mm
        }

\vspace{0.3cm}

\vfill
{\large
 J. Fleischer$^{a}$, F. Jegerlehner$^{b}$ and
  O.V. Tarasov$^{a,b,}$\footnote{On leave of absence from JINR,
141980 Dubna (Moscow Region), Russian Federation.}$^,\!\!$
\footnote{
Supported by DFG under contract FL 241/4-2
}
}

\vspace{0.7cm}
$^a$~~Fakult\"at f\"ur Physik~~~~~\\
Universit\"at Bielefeld \\
Universit\"atsstr. 25\\
D-33615 Bielefeld, Germany\\
\vspace{0.3cm}

$^b$~~Deutsches Elektronen-Synchrotron DESY~~~~~ \\
Platanenallee 6, D--15738 Zeuthen, Germany\\
\end{center}

\vfill

\begin{abstract}
A difference equation w.r.t. space-time dimension
$d$ for $n$-point one-loop integrals with arbitrary momenta and
masses is introduced and a solution presented. The result can in 
general be written as multiple
hypergeometric series with ratios of different Gram determinants
as expansion variables. Detailed considerations  for $2-,3-$ and 
$4-$point functions are given. For the $2-$point function we reproduce a 
known result in terms of the Gauss hypergeometric function $_2F_1$. 
For the $3-$point function an expression in terms of
$_2F_1$ and the Appell hypergeometric function $F_1$ is given.
For the $4-$point function a new representation in terms of
$_2F_1$, $F_1$ and the Lauricella-Saran functions $F_S$ is obtained.
For arbitrary $d=4-2\varepsilon$,  momenta and masses the
$2-,3-$ and $4-$point functions admit a simple one-fold integral 
representation.
This representation will be useful for the calculation of 
contributions from the $\varepsilon-$ expansion needed in higher
orders of perturbation theory.
Physically interesting examples of $3-$ and $4-$point functions occurring 
in Bhabha scattering are investigated.
\end{abstract}

\newpage

%%%%%%%%%%%%%%%%%%%%%%%%%%%%%%%%%%%%%%%%%%%%%%%%%%%%%%%
\section{Introduction}
%%%%%%%%%%%%%%%%%%%%%%%%%%%%%%%%%%%%%%%%%%%%%%%%%%%%%%%
The calculation of radiative corrections in the electroweak Standard
Model (SM) is especially demanding because of the many species of
particles and fields and the large variety of interactions between
them as well as the many mass and energy scales which typically show
up in a high energy scattering process. Practical problems in the
calculation and numerical evaluation of Feynman integrals were
encountered at an early stage of SM calculations beyond the tree
level, at the level of the one--loop integrals already. To a large
extent, known analytic results and techniques have been reviewed,
extended and discussed long time ago in ~\cite{'tHooft:1979xw}. In
addition M. Veltman first had the idea to develop a library of
numerical routines, the program {\tt FormF}~\cite{Veltman:1979ff},
which allowed to calculate the one--loop radiative corrections for $ 2
\rightarrow 2$ scattering processes in a numerically stable way. Unfortunately,
this program was only running in CDC assembler, and used very high
precision (more than 100 digits) to overcome the numerical
instabilities that plagues the calculation of one-loop electroweak
radiative correction. Later G.J. van Oldenborgh implemented the most
relevant one--loop integrals as a normal {\tt Fortran} library {\tt
FF}~\cite{vanOldenborgh:1990wn}, which utilizes alternative algorithms
which run in double precision {\tt Fortran 77}. This program was very
useful for many of the calculations which were needed for precision
physics at LEP. Meanwhile a number of extensions to two-- and more--loops
have been developed. For example, the programs
{\tt SHELL2}~\cite{Fleischer:1992xp} and
{\tt MINCER}~\cite{Larin:1991fz} 
allow to calculate special types of integrals.

More recently we face new problems in view of the requirements for
future colliders. For example, we certainly will need full electroweak
two--loop calculations for $ 2 \rightarrow 2$ fermion processes as well as
full one--loop $2 \rightarrow 4$ fermion processes for precision physics at
a future $e^+e^-$ linear collider like TESLA. In both cases the
existing program libraries need important extensions.

In the first case, for the renormalization of two and more loops, one
needs one--loop integrals to higher order in the $\varepsilon$--expansion
($d=4- 2 \varepsilon$), while in the second case efficient algorithms
are needed for the calculation of $5-$, $6-$ and higher-- point functions,
which at least on the level of the scalar integrals are ultraviolet finite in
$d=4$ dimensions. Usually, one may reduce them to a sum of $4-$ and
lower--point functions, however with ``unphysical'' external
kinematics. While {\tt FF} serves all requirements to calculate $2 \rightarrow
2$ processes at one--loop, it does not work in general if we try to
evaluate the expressions we obtain as a result from the reduction of
higher point functions and we thus have to extend this tool
appropriately.

Another important extension we need stems from the fact that unstable
particles, like the massive gauge bosons $W$, $Z$, the Higgs and the
top quark, show up as intermediate states in scattering matrix
elements. Most interesting is their production near resonance, where
ordinary perturbation theory breaks down and requires to work with
finite width propagators not only for single resonant virtual particle
lines but also within loops. Most obviously this shows up in the
interplay between virtual and real photon emission in infrared
sensitive quantities.

In this paper we propose a new method to deal with one--loop integrals
which allows for extensions in several respects. The algorithm we
propose works for arbitrary space--time dimensions, for complex masses
and momenta and allows to find the appropriate physical domain of the
analytic structure.

The paper is organized as follows: in Sect. 2 we recall the basic 
recursion relation and introduce the relevant kinematical quantities.
Based on this recursion relation a difference equation and
its solution is given in Sect. 3, which quite generally represents
$n-$point functions in terms of series over $(n-1)-$point functions.
An asymptotic method to determine the `boundary term' is discussed.
In the following sections, we give explicit representations: in
Sect. 4 for the $2-$point function, where also the approach itself
is demonstrated in great detail; similarly in Sect. 5 for the $3-$point
function and in Sect.6  for the $4-$point function. Explicit examples 
needed in Bhabha scattering are also presented. For the
boundary term of a $4-$point Bhabha diagram a differential equation
is set up as alternative method for its determination.

\section{General remarks on one--loop functions}
%%%%%%%%%%%%%%%%%%%%%%%%%%%%%%%%%%%%%%%%%%%%%%%%%%%%%%%%%%

Relations between Feynman integrals in different dimensions
are known since quite a long time \cite{Young}.
Recurrence relations for $n$-point one-loop integrals, required for our 
investigation, are given in \cite{fjt}, simplifying and extending the work of
\cite{ovt1}.
We consider  scalar one-loop integrals  depending
on $n-1$ external momenta:
%&&&&&&&&&&&&&&&&&&&&&&&&&&&&&&&&&&&&&&&&&&&&&&&&&&&&&&&&&&&&
\begin{equation}
I_n^{(d)}=
\int \frac{d^d q}{[i \pi^{{d}/{2}}]} \prod_{j=1}^{n}
\frac{1}{c_j^{\nu_j}},
\end{equation}
%&&&&&&&&&&&&&&&&&&&&&&&&&&&&&&&&&&&&&&&&&&&&&&&&&&&&&&&&&&&&
where
\begin{equation}
c_j=(q-p_j)^2-m_j^2+i\epsilon.
\end{equation}
The corresponding diagram and the convention for
the momenta are given in Fig.1.

\begin{center}
\vspace*{-2cm}
\vbox{
 \raisebox{5.0cm}{\makebox[0pt]{\hspace*{-2cm}$$}}
 \epsfysize=30mm \epsfbox{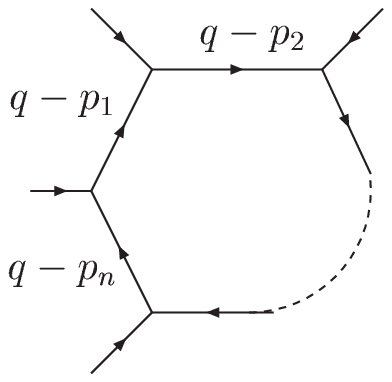}
     }
\end{center}
\vspace*{0.7cm}
\noindent
\begin{center}
Fig. 1: One-loop diagram with $n$ external legs.
\end{center}

A recurrence relation connecting integrals $I^{(d)}_n$ with different  
space-time  dimensions was given in \cite{fjt} 

\begin{equation}
  (d-\sum_{i=1}^{n}\nu_i+1)G_{n-1}I^{(d+2)}_n=
  \left[2 \Delta_n+\sum_{k=1}^n (\partial_k \Delta_n) {\bf k^-}
  \right]I^{(d)}_n,
  \label{reduceDtod}
\end{equation}

where $\partial_j \equiv \partial / \partial m_j^2,$ and
\vspace{0.8cm}

\begin{equation}
\Delta_n \equiv \Delta_n(\{p_1,m_1\},\ldots \{p_n,m_n\})=  \left|
\begin{array}{cccc}
Y_{11}  & Y_{12}  &\ldots & Y_{1n} \\
Y_{12}  & Y_{22}  &\ldots & Y_{2n} \\
\vdots  & \vdots  &\ddots & \vdots \\
Y_{1n}  & Y_{2n}  &\ldots & Y_{nn}
\end{array}
         \right|,~~~~~~~~~~~~~~~~~~~~~~
\label{deltan}
\end{equation}
\begin{equation}
Y_{ij}=-(p_i-p_j)^2+m_i^2+m_j^2,
\end{equation}

\begin{eqnarray}
&&G_{n-1} \equiv G_{n-1}(p_1,\ldots ,p_n)= \nonumber \\
&& \nonumber \\
&&~~-2^n \left|
\begin{array}{cccc}
  \!(p_1-p_n)(p_1-p_n)  & (p_1-p_n)(p_2-p_n)  &\ldots & 
  (p_1-p_n)(p_{n-1}-p_n) \\
  \!(p_1-p_n)(p_2-p_n)  & (p_2-p_n)(p_2-p_n) &\ldots & 
  (p_2-p_n)(p_{n-1}-p_n) \\
  \vdots  & \vdots  &\ddots & \vdots \\
  \!(p_1-p_n)(p_{n-1}-p_n)  & (p_2-p_n)(p_{n-1}-p_n) 
  &\ldots & (p_{n-1}-p_n)(p_{n-1}-p_n)
\end{array}
\right|, ~~~~
\label{Gn}
\end{eqnarray}
$$G_0 \equiv - 2,$$
$p_i$ are combinations of external momenta flowing through $i$-th
lines, respectively, and $m_i$ is the mass of the $i$-th line.
In the following we also quite often use the abbreviation 
$p_{ij}^2=(p_i-p_j)^2$.
By shifting the integration momentum we choose in general $p_n=0$.
Where no confusion can arise, we simply refer to the above
functions as $\Delta_n$, $G_{n-1}$. Considering integrals for
$n=2,3,4$ in the next
sections we will use also an indexed notation for $\Delta_n$ and $G_{n-1}$
%&&&&&&&&&&&&&&&&&&&&&&&&&&&&&&&&&&&&&&&&&&&&&&&&&&&&&&&&&&&&
\begin{eqnarray}
&&\lambda_{i_1 i_2 \ldots i_n } = 
 \Delta_n(\{p_{i_1},m_{i_1}\},\{p_{i_2},m_{i_2} \},
  \ldots ,\{p_{i_n},m_{i_n}\}),                         \nonumber \\
&& \nonumber \\
&&g_{i_1 i_2 \ldots i_n }=G_{n-1}(p_{i_1},p_{i_2},\ldots ,p_{i_n}).
\label{lage}
\end{eqnarray}
%&&&&&&&&&&&&&&&&&&&&&&&&&&&&&&&&&&&&&&&&&&&&&&&&&&&&&&&&&&&&
The indexed notation will be useful when considering integrals
obtained from $I_n^{(d)}$ by contracting some lines.
Rather frequently the results will depend on ratios of 
$\lambda_{i_1 i_2 \ldots i_n}$  and $g_{i_1 i_2 \ldots i_n}$  and therefore 
it is  convenient to introduce the notation
\begin{equation}
r_{ij\ldots k}=-\frac{\lambda_{ij\ldots k}}{g_{ij\ldots k}}.
\end{equation}

With this definition a useful relation is:
\begin{equation}
r_{i_1{\ldots} i_{k-1}i_k i_{k+1} {\ldots} i_n}
-r_{i_1{\ldots} i_{k-1} i_{k+1} {\ldots} i_n}
=- \frac{( \partial_{i_k} \lambda_{i_1 {\ldots}  i_n})^2}
{ 2 g_{i_1 {\ldots}  i_n}~  g_{i_1 {\ldots}  i_{k-1}i_{k+1}{\ldots} i_n}}.
\end{equation}

Using 
\begin{equation}
\sum_{j=1}^n \partial_j \lambda_{i_1{\ldots} i_n} = -g_{i_1{\ldots} i_n}= -G_{n-1}
\label{Gnm1}
\end{equation}
one shows that to all orders in $\epsilon$
\begin{equation}
\lambda_{i_1 i_2 \ldots  i_n}(\{m_r^2-i\epsilon\})=
\lambda_{i_1 i_2 \ldots  i_n}(\{m_r^2\})+i g_{i_1 i_2 \ldots i_n} ~ \epsilon,
\end{equation}
and therefore the $\epsilon$ prescription for $r$ is rather simple
(with the same $\epsilon$ for all masses)
\begin{equation}
\left. r_{ij\ldots k}\right|_{m_j^2-i\epsilon}=
\left. r_{ij\ldots k}\right|_{m_j^2}-i\epsilon.
\label{repsilon}
\end{equation}
As we will see later, the results for one loop integrals will
be expressed in terms of hypergeometric functions depending
on ratios of different $r$'s plus terms proportional to
$r_{i{\ldots} j}^{d/2}$. Relation (\ref{repsilon}) makes the $r$'s
appear like masses, i.e. their $i\epsilon$ prescription is the
same as for masses (their dimension is mass squared). This 
simple property is of extreme importance
for the study of the analytic behavior
of one loop integrals, in particular their analytic continuation 
to different kinematical regions for arbitrary space-time dimension.
For brevity, we shall omit ``causal'' $i\epsilon$'s below.

%%%%%%%%%%%%%%%%%%%%%%%%%%%%%%%%%%%%%%%%%%%%%%%%%%%%%%%%%%%%%%%%
%%%%%%%%%%%%%%%%%%%%%%%%%%%%%%%%%%%%%%%%%%%%%%%%%%%%%%%%%%%%%%%%
\section{Solution of the difference equation}
%%%%%%%%%%%%%%%%%%%%%%%%%%%%%%%%%%%%%%%%%%%%%%%%%%%%%%%%%%%%%%%%
%%%%%%%%%%%%%%%%%%%%%%%%%%%%%%%%%%%%%%%%%%%%%%%%%%%%%%%%%%%%%%%%

In this section we will derive an explicit solution of relation
(\ref{reduceDtod})  for $\nu_i=1$.  This  relation  has the
simpler form
%&&&&&&&&&&&&&&&&&&&&&&&&&&&&&&&&&&&&&&&&&&&&&&&&&&&&&&&&&&&&&&
\begin{equation}
  (d-n+1)G_{n-1}I^{(d+2)}_n= \left[2 \Delta_n+\sum_{k=1}^n 
  (\partial_k \Delta_n) {\bf k^-}\right]I^{(d)}_n.
  \label{nuieq1}
\end{equation}
%&&&&&&&&&&&&&&&&&&&&&&&&&&&&&&&&&&&&&&&&&&&&&&&&&&&&&&&&&&&&&&
If we assume that, evaluating $n$ -point functions, we already know
$n-1$ point functions, then relation (\ref{nuieq1}) represents an
inhomogeneous first order difference equation with respect to $d$. 
Methods to 
solve this kind of equations are well described in the mathematical 
literature \cite{Miller}, \cite{Milne}. By the redefinition 
%&&&&&&&&&&&&&&&&&&&&&&&&&&&&&&&&&&&&&&&&&&&&&&&&&&&&&&&&&&&&&&&&&&&
\begin{equation}
   I_{n}^{(d)}=\frac{1}{\Gamma\left(\frac{d-n+1}{2}\right)}
   \left( \frac{\Delta_n}{G_{n-1}}\right)^{\frac{d}{2}}
   \overline{I}_{n}^{(d)}
\label{defIbar}
\end{equation}
%&&&&&&&&&&&&&&&&&&&&&&&&&&&&&&&&&&&&&&&&&&&&&&&&&&&&&&&&&&&&&&&&&&&
we obtain the simpler equation  
%&&&&&&&&&&&&&&&&&&&&&&&&&&&&&&&&&&&&&&&&&&&&&&&&&&&&&&&&&&&&&&
\begin{equation}
   \overline{I}_{n}^{(d+2)}=\overline{I}_{n}^{(d)}
   +\frac{\Gamma \left(\frac{d-n+1}{2}\right)}
   {2 \Delta_n} \left(\frac{G_{n-1}}{\Delta_n}\right)^{\frac{d}{2}}
   \sum_{k=1}^{n}(\partial_k \Delta_n) {\bf k^{-}} I_{n}^{(d)}.
\label{Ibarequ}   
\end{equation}
%&&&&&&&&&&&&&&&&&&&&&&&&&&&&&&&&&&&&&&&&&&&&&&&&&&&&&&&&&&&&&&
Without loss of generality we can parameterize $d$ as
\begin{equation}
d=2l - 2 \varepsilon,
\end{equation}
where $l$ is integer and $\varepsilon$ small and possibly complex.
Then the solution of the equation for $\overline{I}_{n}^{(d)}$ can be
written as 
%&&&&&&&&&&&&&&&&&&&&&&&&&&&&&&&&&&&&&&&&&&&&&&&&&&&&&&&&&&&&&&&&&&
\begin{equation}
  \overline{I}_{n}^{(2l-2\varepsilon)}=\sum_{r=0}^{l}
  \frac{\Gamma\left(r-1-\varepsilon-\frac{n-1}{2}\right)}
  {2\Delta_n}\left(\frac{G_{n-1}}{\Delta_n}\right)^{r-1-\varepsilon}
  \sum_{k=1}^{n}(\partial_k \Delta_n) {\bf k^{-}}I_{n}^{(2r-2
  -2\varepsilon)}+\tilde{b}_n(\varepsilon),
\label{solution1}
\end{equation}
%&&&&&&&&&&&&&&&&&&&&&&&&&&&&&&&&&&&&&&&&&&&&&&&&&&&&&&&&&&&&&&&&&
where $\tilde{b}_n(\varepsilon)=\overline{I}_{n}^{(-2-2\varepsilon)}$ 
is an $l$ independent constant. 
With $\sum_{r=0}^{l}=\sum_{r=0}^{\infty}-\sum_{r=l+1}^{\infty}$
and shifting the summation index in the second sum like $r \rightarrow r+l+1$, 
the solution (\ref{solution1}) can be rewritten in the form
%&&&&&&&&&&&&&&&&&&&&&&&&&&&&&&&&&&&&&&&&&&&&&&&&&&&&&&&&&&&&&&&&&
\begin{equation}
   \overline{I}_{n}^{(2l-2\varepsilon)} = -\sum_{r=0}^{\infty}
   \frac{\Gamma\left(r+\frac{d-n+1}{2} \right)}{2\Delta_n}
   \left(\frac{G_{n-1}}{\Delta_n}\right)^{r+\frac{d}{2}}
   \sum_{k=1}^{n}(\partial_k \Delta_n) {\bf k^{-}}I_{n}^{(2r+d)}
   +\overline{b}_n(\varepsilon)
\label{solution2}
\end{equation}
by redefining the $l$ independent `boundary term' $\tilde{b}_n$.
The final result for $I_n^{(d)}$ then reads
\begin{equation}
I_n^{(d)}=b_n(\varepsilon)
  -\sum_{k=1}^{n} \left(\frac{\partial_k \Delta_n}
  {2 \Delta_n}\right)\sum_{r=0}^{\infty} \left(\frac{d-n+1}{2}
  \right)_r\left(\frac{G_{n-1}}{\Delta_n}\right)^{r}
  {\bf k^{-} } I^{(d+2r)}_n.
\label{npoint}
\end{equation}
As we will show in the next sections, $b_n$ can be determined
from the asymptotic behavior of $I_n^{(d)}$ for $d \rightarrow \infty$ 
or by setting up a differential equation for it. This term depends 
on the kinematic domain.

%&&&&&&&&&&&&&&&&&&&&&&&&&&&&&&&&&&&&&&&&&&&&&&&&&&&&&&&&&&&&&&
The series in the above solution converge in general if the expansion parameter
does not exceed 1. If it does, one has
to continue the series analytically. This can be done by different
methods. If the result is already obtained in terms of known
hypergeometric functions, e.g., then known formulae for their analytic 
continuation can be applied.
Another method will be to modify the procedure of solving
the difference equation. According to the general theory of
difference equations, we should repeat the derivation in our case 
by using the following parameterization for $d$ (changing the sign of $l$):

\begin{equation}
d= - 2l - 2 \varepsilon.
\end{equation}

With this parameterization  for the function 

\begin{equation}
f_l = \overline{I}^{(-2l-2\varepsilon)},
\end{equation}

we obtain from Eq. (\ref{Ibarequ})
%&&&&&&&&&&&&&&&&&&&&&&&&&&&&&&&&&&&&&&&&&&&&&&&&&&&&&&&&&&&&&&
\begin{equation}
  f_{l+1}= f_l
   -\frac{\Gamma \left(\frac{-2l-2\varepsilon-n+1}{2}\right)}
   {2 \Delta_n} \left(\frac{G_{n-1}}{\Delta_n}\right)^
   {-l-\varepsilon -1}
   \sum_{k=1}^{n}(\partial_k \Delta_n) {\bf k^{-}} I_{n}^
   {(-2l-2-2\varepsilon)}.
\label{flequ}   
\end{equation}
%&&&&&&&&&&&&&&&&&&&&&&&&&&&&&&&&&&&&&&&&&&&&&&&&&&&&&&&&&&&&&&

Solving   this equation we obtain for $I_{n}^{(d)}$
%&&&&&&&&&&&&&&&&&&&&&&&&&&&&&&&&&&&&&&&&&&&&&&&&&&&&&&&&&&&&&&&&&&
\begin{equation}
I_{n}^{(d)}=b_n(\varepsilon) -\frac{1}{(d-n-1)}  \nonumber \\
 \sum_{k=1}^{n} \left(\frac{\partial_k \Delta_n}{\Delta_n} \right)
\sum_{r=0}^{\infty} \frac{1}{\left(\frac{3+n-d}{2}\right)_r}
  \left(-\frac{\Delta_n}{G_{n-1}}\right)^{r+1}
   {\bf k^{-}}I_{n}^{(d-2r-2)}.
\label{solution1a}
\end{equation}
where $(a)_r \equiv \Gamma(r+a)/\Gamma(a)$ is the Pochhammer symbol.
Here we have for convenience introduced the same notation for the boundary
term as before, keeping in mind, however, that the value may be different 
from the one in (\ref{npoint}).
In order to obtain convergent series for all the contributions in $\sum_k^n$,
one can use different parameterizations for $d$ in the various terms.

   In general the situation is such that to obtain the multiple series
is straightforward. The problem is rather to determine the boundary term.
In the next chapter we discuss for this purpose the asymptotic 
behaviour for $d \rightarrow \infty$.

%%%%%%%%%%%%%%%%%%%%%%%%%%%%%%%%%%%%%%%%%%%%%%%%%%%%%%%%%%%%%%%%
%%%%%%%%%%%%%%%%%%%%%%%%%%%%%%%%%%%%%%%%%%%%%%%%%%%%%%%%%%%%%%%%
\subsection{Asymptotic behavior of $I_n^{(d)}$ for $d \rightarrow \infty$}
%%%%%%%%%%%%%%%%%%%%%%%%%%%%%%%%%%%%%%%%%%%%%%%%%%%%%%%%%%%%%%%%
%%%%%%%%%%%%%%%%%%%%%%%%%%%%%%%%%%%%%%%%%%%%%%%%%%%%%%%%%%%%%%%%

For arbitrary $n$ the asymptotic value of $I_{n}^{(d)}$ as 
$d \rightarrow \infty$ can be found by utilizing the parametric 
representation for $I_n^{(d)}$, i.e., with
%&&&&&&&&&&&&&&&&&&&&&&&&&&&&&&&&&&&&&&&&&&&&&&&&&&&&&&&&&&&&&
\begin{equation}
    \frac{1}{c_1 c_2 \ldots c_n}=
    \int_{0}^{1} \! \ldots \int_0^1 
    \frac{dx_1  \ldots dx_{n-1}~~~~\Gamma(n)~~
    x_1^{n-2} x_2^{n-3} \ldots x_{n-2}}
    {[c_1 x_1\ldots x_{n-1}\!\!+c_2x_1\ldots x_{n-2}
    (1\!-\!x_{n-1})\!+\!\ldots \!+\!c_n(1\!-\!x_1)]^n},
\end{equation}
%&&&&&&&&&&&&&&&&&&&&&&&&&&&&&&&&&&&&&&&&&&&&&&&&&&&&&&&&&&&&&
shifting $q$ in order to remove the linear term and integrating 
over $q$ by means of
%&&&&&&&&&&&&&&&&&&&&&&&&&&&&&&&&&&&&&&&&&&&&&&&&&&&&&&&&&&&&&
\begin{equation}
   \int \frac{d^dq}{[i \pi^{d/2}]}\frac{1}{{(q^2-m_i^2)}^{\alpha}}
   =(-1)^{\alpha} \frac{\Gamma\left(\alpha-\frac{d}{2}\right)}
   {\Gamma(\alpha) (m_i^2)^{\alpha-\frac{d}{2}}},
\label{tadpole}
\end{equation}
%&&&&&&&&&&&&&&&&&&&&&&&&&&&&&&&&&&&&&&&&&&&&&&&&&&&&&&&&&&&&&
any $n$-point function can be represented as a multiple
parametric integral of the form:
%&&&&&&&&&&&&&&&&&&&&&&&&&&&&&&&&&&&&&&&&&&&&&&&&&&&&&&&&&&&&&
\begin{equation}
   I_n^{(d)}\!=\Gamma\left(n\!-\frac{d}{2}\right)
   \int_0^1 \!\!\ldots \!\int_0^1\! dx_1\ldots dx_{n-1} 
   f(x_1,\ldots,x_{n-2})
   (h_n(\{p_j,m_s\},x_1,\ldots,x_{n-1}))^{\frac{d}{2}-n}.
\end{equation}
%***********************************************************

In the analyticity domain of $I_n^{(d)}$ the polynomial
$h_n(\{p_j,m_s\},x_1,{\ldots} ,x_{n-1})
 \geq 0$ and $I_n^{(d)}$ is an integral of Laplace type.
In the domain of nonanalyticity there are subdomains of the 
integration region where $h_n(\{p_j,m_s\},x_1,{\ldots} ,x_{n-1}) < 0$  
and due to that the integral $I_n^{(d)}$ gets an imaginary part. 
For this kinematical configuration $I_n^{(d)}$ may be represented as
\begin{eqnarray} 
&&   I_n^{(d)}\!=\Gamma\left(n\!-\frac{d}{2}\right)\left \{
   \int  \{dx\}   \theta(h_n) f(\{x\})
   (h_n(\{p_j,m_s\},\{x\}))^{\frac{d}{2}-n}\right.
   \nonumber \\
&& \nonumber \\
&&~~ +\cos \frac{\pi}{2}(d-2n) 
 \sum_j \int_{\Omega^-_j} \{dx\} f(x_1,\ldots,x_{n-2})
   \left| h_n(\{p_j,m_s\},x_1,\ldots,x_{n-1})\right|^{\frac{d}{2}-n}    
 \nonumber \\
&&~~\left. \pm i \sin \frac{\pi}{2}(d-2n)
   \sum_j \int_{\Omega^-_j} \{dx\} f(x_1,\ldots,x_{n-2})
   \left| h_n(\{p_j,m_s\},x_1,\ldots,x_{n-1})\right|
        ^{\frac{d}{2}-n}    
  \right \}.
\label{RePlusIm}  
\end{eqnarray}
Here $\Omega^-_j$ are  subdomains of the integration region where $h_n<0$.
The sign of the imaginary part has to be determined from the $i \epsilon$
prescription in $h_n$.

Our parametric integrals are of multiple Laplace type which
in general can be written as
%&&&&&&&&&&&&&&&&&&&&&&&&&&&&&&&&&&&&&&&&&&&&&&&&&&&&&&&&&&&&&&&&&
\begin{equation}
    F(\lambda)=\int_{\Omega}f(x) \exp[\lambda S(x)]dx
\end{equation}
%&&&&&&&&&&&&&&&&&&&&&&&&&&&&&&&&&&&&&&&&&&&&&&&&&&&&&&&&&&&&&&&&&
(identifying $S(x) \equiv ln(\left| h(x) \right|)$ and $\lambda \equiv
 \frac{d}{2}-n$ ),
where $\Omega$ is a bounded simply connected domain in $k$ dimensional
Euclidean space, $x=(x_1,\ldots x_k)$ and $S(x)$ is a real function. 
The functions $S(x)$ and $f(x)$ are sufficiently
differentiable functions of their arguments throughout $\Omega$.
Supposing that a relative maximum of $S(x)$ in $\Omega$ is achieved
at the interior point $x=\overline{x}$ ,
\footnote{
If the function $h_n$ reaches its maximum value on the boundary of the
integration region then \\
%\begin{equation}
$   F(\lambda) \sim \exp[\lambda S(\overline{x})] 
   \sum_{r=0}^{\infty} a_r \lambda^{-r-\frac{k+1}{2}}$.
}
 then at $\lambda \rightarrow \infty$  \cite{Fedoryuk}
%&&&&&&&&&&&&&&&&&&&&&&&&&&&&&&&&&&&&&&&&&&&&&&&&&&&&&&&&&&&&&&&&
\begin{equation}
   F(\lambda) \sim \exp[\lambda S(\overline{x})] 
   \sum_{r=0}^{\infty} a_r \lambda^{-r-\frac{k}{2}}
 \label{asym}
\end{equation}
%&&&&&&&&&&&&&&&&&&&&&&&&&&&&&&&&&&&&&&&&&&&&&&&&&&&&&&&&&&&&&&&&
(correspondingly this holds for a relative minimum).
This expansion may be differentiated w.r.t. $\lambda$ any
number of times. The leading term of the expansion is
%&&&&&&&&&&&&&&&&&&&&&&&&&&&&&&&&&&&&&&&&&&&&&&&&&&&&&&&&&&&&&&&&&
\begin{equation}
   F(\lambda)=\exp[\lambda S(\overline{x})]~ (2\pi/\lambda)^{\frac{k}
   {2}}~\frac{f(\overline{x})+O(\lambda^{-1})}{\sqrt{\left|{\rm det}
   S_{xx}(\overline{x})\right|}},
\label{maxinside} 
\end{equation}
%&&&&&&&&&&&&&&&&&&&&&&&&&&&&&&&&&&&&&&&&&&&&&&&&&&&&&&&&&&&&&&&&&
where ${\rm det}S_{xx}$ is the Hessian determinant, the indices 
appearing as second derivatives.
The behavior of $I_n^{(d)}$ for $d \rightarrow \infty$
can thus be obtained. An extremum is found from
\begin{equation}
\frac{\partial h_n}{\partial x_i}=0,
\label{extr}
\end{equation}
and using the determinant representation of $h_n$, 
it is possible to show that
\footnote{There are other solutions, but for them in general
one of the $\overline{x}_i (i=1, \dots n-2)$ equals $0$ such that
in these cases no contribution to $I_n^{(d)}$ is obtained}
\begin{equation}
\overline{x}_i=\frac{\sum_{k=1}^{n-i} \partial_k \Delta_n}
{\sum_{j=1}^{n-i+1} \partial_j \Delta_n},
\label{xbar}
\end{equation}
and at this point
\begin{equation}
 h_n(\{p_j,m_s\},\overline{x}_1,{\ldots} ,\overline{x}_{n-1})=
 r_{1{\ldots} n}.
 \label{hnrn}
\end{equation}
All $\overline{x}_i$ are inside the interior region, i.e.
\begin{equation}
0 \leq \overline{x}_i \leq 1,~~~ (i=1, \dots ,n-1)
\label{xbarli}
\end{equation}
if all derivatives $\partial_k \Delta_n$ have the same sign.
Conversely, it is easy to see that, provided (\ref {xbarli}) is
true, due to (\ref {Gnm1}) all $\partial_k \Delta_n$ have
the opposite sign as $G_{n-1}$: assuming 
$\sum_{k=1}^n \partial_k \Delta_n \ge 0 (\le 0)$, from the left hand 
side of the inequalities (\ref{xbarli}) follows
$\sum_{k=1}^{n-i} \partial_k \Delta_n \ge 0 (\le 0), i=1, \dots ,n-1$.
Multiplying (\ref {xbarli}) with the positive (negative) denominators
of $\overline{x}_i$, the right hand side yields 
$\partial_{n-i+1} \Delta_n \ge 0 (\le 0), i=1, \dots ,n-1$.

The general idea to determine the boundary term $b_n$ from the 
asymptotic behavior as $d \rightarrow \infty$ is as follows:
from (\ref{defIbar}) and (\ref{solution2}) we see that for large
$d$
\begin{equation}
b_n \sim  r_{1 \dots n}^{d/2}.
\end{equation}
Such a contribution can come from the asymptotic behavior of
$I_n^{(d)}$ on the l.h.s. of (\ref{npoint}) due to (\ref{asym}) 
and (\ref{hnrn}). This may happen, however, only if all 
$0< \overline{x}_i<1$, i.e. if for the specific kinematics under
consideration an extremum occurs inside the integration region
and an absolute maximum of $|h_n|$ does not occur on the border.
Finally we can write in the region where the integrand of $I^{(d)}_n$ 
has an extremum inside the integration region
%&&&&&&&&&&&&&&&&&&&&&&&&&&&&&&&&&&&&&&&&&&&&&&&&&&&&&&&&&&&&&&&&&
\begin{eqnarray}
&& I_n^{(d)}=-(2 \pi)^{\frac{n}{2}}
  \frac{\Gamma\left(1-\frac{d}{2}\right) \Gamma \left(
  \frac{d}{2}
  \right)}{ \Gamma \left(\frac{d-n+1}{2}\right)}
  \frac{r_n^{\frac{d-n-1}{2}  }}
  {\sqrt{ \pi |G_{n-1}|}}
\nonumber \\  
&&~~~~~~~~~~~~~~~~~~~~~~~~~
-\sum_{k=1}^{n} \left(\frac{\partial_k \Delta_n}
  {2 \Delta_n}\right)\sum_{r=0}^{\infty} \left(\frac{d-n+1}{2}
  \right)_r    \times \left(\frac{G_{n-1}}{\Delta_n}\right)^{r}
  {\bf k^{-} } I^{(d+2r)}_n,
\label{npointEucl}
\end{eqnarray}
%&&&&&&&&&&&&&&&&&&&&&&&&&&&&&&&&&&&&&&&&&&&&&&&&&&&&&&&&&&&&&&&&
where $r_n=r_{1 \dots n}$ 
\footnote{
If the function $h_n$ reaches its absolute maximum value on the boundary 
of the integration region then $I^{(d)}_n$
has no asymptotic behavior like $\sim  r_{1 \dots n}^{d/2}$ and $b_n=0$.}.
To determine the boundary term, convergent series are requested
on the r.h.s. of (\ref {npointEucl}). These series are convergent 
in general only in certain kinematical domains and only for these
is the obtained boundary term the correct one. The method of analytic
continuation to other domains has been indicated above. 
In fact, the summation with $l > 0$ and $l < 0$, respectively, 
for the analytic continuation turns out to be easier in general than 
the direct analytic continuation of hypergeometric functions.
In the next section we demonstrate the 
procedure in all details for the $2-$point function and also give
an example for the summation with $l < 0$.

Analytic continuation of the (generalized) hypergeometric functions
can also be useful to find the asymptotic behavior, like double
logarithms, for certain diagrams in specific kinematic regions.

%%%%%%%%%%%%%%%%%%%%%%%%%%%%%%%%%%%%%%%%%%%%%%%%%%%%%%%%%%%%%%%%%
%%%%%%%%%%%%%%%%%%%%%%%%%%%%%%%%%%%%%%%%%%%%%%%%%%%%%%%%%%%%%%%%%
\section{2-point function}
%%%%%%%%%%%%%%%%%%%%%%%%%%%%%%%%%%%%%%%%%%%%%%%%%%%%%%%%%%%%%%%%%
%%%%%%%%%%%%%%%%%%%%%%%%%%%%%%%%%%%%%%%%%%%%%%%%%%%%%%%%%%%%%%%%%

We find it convenient to label the lines of $I_2^{(d)}$ as $i,j$
because the integrals $I_2^{(d)}$ will be encountered in calculating 
$I_3^{(d)}$ (and $I_4^{(d)}$) as a result of contraction of different 
lines.
Expression (\ref{npoint}) for $I_2^{(d)}$ includes two one-fold
sums over tadpole integrals $I_1^{(d)}$, 
given in (\ref{tadpole}). $I_1^{(d)}$ can also be obtained
from (\ref{npoint}) assuming  that in dimensional regularization 
$I_0^{(d)}=0$
%&&&&&&&&&&&&&&&&&&&&&&&&&&&&&&&&&&&&&&&&&&&&&&&&&&&&&&&&&&&&&
\begin{equation}
  I_1^{(d)}(m_i)=-\Gamma\left(1-\frac{d}{2}\right)
  (m_i^2)^{\frac{d-2}{2}}.
\label{I1}
\end{equation}
%&&&&&&&&&&&&&&&&&&&&&&&&&&&&&&&&&&&&&&&&&&&&&&&&&&&&&&&&&&&&&

For $n=2$, substituting (\ref{I1}) into (\ref{npoint}) gives 
(observe the different normalization of $b_2$ in comparison 
with (\ref{npoint})):
\begin{eqnarray}
\frac{2 \lambda_{ij} I_2^{(d)} }{\Gamma\left(1-\frac{d}{2}\right)} =
 b_2&+& \frac{\partial_i \lambda_{ij}}{(m_j^2)^{1-\frac{d}{2}}}
   \sum_{r=0}^{\infty} \frac{\left( \frac{d-1}{2}
   \right)_r}{\left(\frac{d}{2}\right)_r}
   \left(- \frac{m_j^2 G_1}{\lambda_{ij}}\right)^r 
 \nonumber \\
&& \nonumber \\
~~ 
   &+&\frac{\partial_j \lambda_{ij}}{(m_i^2)^{1-\frac{d}{2}}}
    \sum_{r=0}^{\infty} \frac{ \left(\frac{d-1}{2}\right)_r}
   {\left(\frac{d}{2}\right)_r}\left(-\frac{m_i^2 G_1}{\lambda_{ij}}
   \right)^r,
\label{i2sums}
\end{eqnarray}
where
\begin{equation}
G_1 = -4p_{ij}^2.
\end{equation}
and the infinite series in (\ref{i2sums}) can be represented
as hypergeometric functions, i.e.
\begin{equation}
 \sum_{r=0}^{\infty} \frac{ \left(\frac{d-1}{2}\right)_r}
   {\left(\frac{d}{2}\right)_r} z^r = \Fh21\Ffz{1,\frac{d-1}{2}}{\frac{d}{2}}
\label{2F1}
\end{equation}
The parametric formula for $I_2^{(d)}$ is a one-fold
integral:
%&&&&&&&&&&&&&&&&&&&&&&&&&&&&&&&&&&&&&&&&&&&&&&&&&&&&&&&&&&&&&
\begin{equation}
  I_2^{(d)}=\Gamma \left(2-\frac{d}{2}\right)\int_0^1 
  dx_1~[p_{ij}^2x_1^2-x_1(p_{ij}^2-m_i^2+m_j^2)
  +m_j^2]^{\frac{d}{2}-2}.
\end{equation}
%&&&&&&&&&&&&&&&&&&&&&&&&&&&&&&&&&&&&&&&&&&&&&&&&&&&&&&&&&&&&&
The extremum of $h_2$ is located at
%&&&&&&&&&&&&&&&&&&&&&&&&&&&&&&&&&&&&&&&&&&&&&&&&&&&&&&&&&&&&&
\begin{equation}
   \overline{x}_1=\frac{\partial_1 \lambda_{ij}}{\partial_1 \lambda_{ij} 
   + \partial_2 \lambda_{ij}}=
   \frac{p_{ij}^2-m_i^2+m_j^2}{2p_{ij}^2},
 \label{x1}
\end{equation}
%&&&&&&&&&&&&&&&&&&&&&&&&&&&&&&&&&&&&&&&&&&&&&&&&&&&&&&&&&&&&&
where
\begin{equation}
  \lambda_{ij}=-(p_{ij}^2)^2-m_i^4-m_j^4+2p_{ij}^2m_i^2
  +2p_{ij}^2m_j^2+2m_i^2m_j^2.
\end{equation}
and 
\begin{equation}
\partial_i \lambda_{ij}=2(p_{ij}^2-m_i^2+m_j^2)~~,
~~\partial_j \lambda_{ij}=2(p_{ij}^2+m_i^2-m_j^2).
\end{equation}
Since
%&&&&&&&&&&&&&&&&&&&&&&&&&&&&&&&&&&&&&&&&&&&&&&&&&&&&&&&&&&&&&
\begin{equation}
  \frac{\partial^2 h_2}{\partial x_1^2}=2p_{ij}^2,
\end{equation}
%&&&&&&&&&&&&&&&&&&&&&&&&&&&&&&&&&&&&&&&&&&&&&&&&&&&&&&&&&&&&&
a maximum of $h_2$ inside the integration region exists for Euclidean
momenta $p_{ij}^2<-(m_j^2-m_i^2)$ (without loss of generality we assume 
here $m_j \ge m_i$). A minimum exists `inside' for $p_{ij}^2>+(m_j^2-m_i^2)$.
In the first case we have $\partial_1 \lambda_{ij} < 0$ and 
$\partial_2 \lambda_{ij} <0$ while in the second case we  have 
$\partial_1 \lambda_{ij} > 0$ and $\partial_2 \lambda_{ij} >0$, 
i.e. in both cases the two derivatives have the same sign.

\begin{center}
\vspace*{0.1cm}
\vbox{
 \raisebox{5.0cm}{\makebox[0pt]{\hspace*{1cm}$$}}
 \epsfysize=80mm \epsfbox{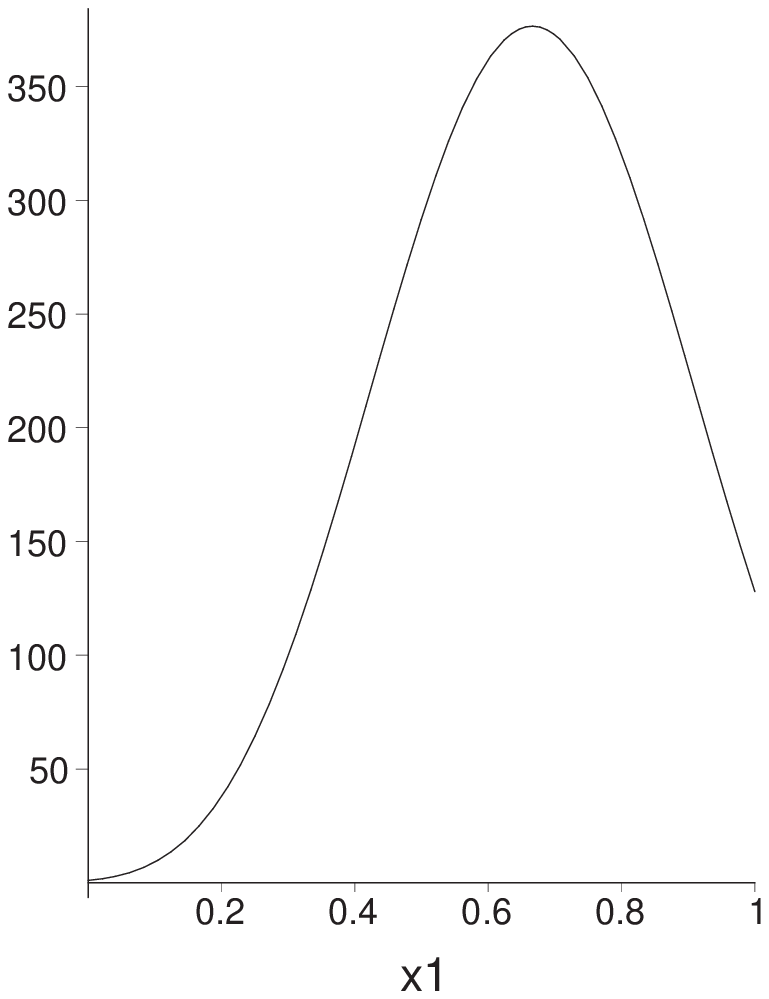}
     }
\end{center}
\vspace*{-0.5cm}
\noindent
\begin{center}
Fig. 2: Function $h_2^{d/2-2}$ at $p_{ij}^2=-3$, $m_i^2=2,m_j^2=1$,
 $d=18$.
\end{center}

For completeness we also give the imaginary part of $I_2^{(d)}$
on the cut, which we obtain from (\ref{RePlusIm}):
\begin{equation}
{\rm Im}~I_2^{(d)}=\pm \Gamma \left(2-\frac{d}{2}\right)
 \sin \frac{\pi(d-2)}{2} (-p_{ij}^2)^{(\frac{d}{2}-2)} 
 \int^{x^+_1}_{x^-_1} d x_1 ( (x_1-x^-_1)(x^+_1 -x_1))^{\frac{d}{2}-2},
 \label{Ima2}
\end{equation}
where
\begin{equation}
x^{\pm}_1=\frac{ p_{ij}^2-m_i^2+m_j^2 \pm \sqrt{-\lambda_{ij}}}
 {2p_{ij}^2}
\end{equation} 
are solutions of the equation
\begin{equation}
h_2(\{p_{ij},m_i,m_j\},x_1)=0.
\end{equation}
For $p_{ij}^2>(m_i+m_j)^2$ both $0 \leq x^{\pm}_1 \leq 1$.
Performing the integration in (\ref{Ima2}), we obtain
\begin{equation}
{\rm Im}~I_2^{(d)}= \frac{\mp \pi \Gamma \left( \frac{d-2}{2} \right)}
                            {p_{ij}^{4-d} \Gamma \left( d-2 \right)}
                \left( \frac{ \sqrt{-\lambda_{ij}}}{p_{ij}^2}\right)^{d-3}
\end{equation}
in agreement with the result \cite{Scharf:1993ds} obtained by another method.
Determining now $b_2$, we have to investigate the asymptotic
behavior of $I_2^{(d)}$ for large $d$. For convenience we
give the following expressions for $r_{ij}$:
\begin{equation}
r_{ij}=-\frac{\left[p_{ij}^2 -(m_i+m_j)^2 \right] 
              \left[p_{ij}^2 -(m_i-m_j)^2 \right]}{4 p_{ij}^2}
\label{rij}
\end{equation}
and 
\begin{equation}
r_{ij}-m_j^2=-\frac{(\partial_i \lambda_{ij})^2}{16 p_{ij}^2}, \nonumber
r_{ij}-m_i^2=-\frac{(\partial_j \lambda_{ij})^2}{16 p_{ij}^2}.
\end{equation}
We begin with 
$p_{ij}^2<-(m^2_j-m^2_i) < 0$. Here we have a maximum like the one shown in 
Fig. 2 with $0 < \overline{x}_1 < 1$.  In this case $r_{ij} > m_i^2,m_j^2$ 
so that the hypergeometric
series in (\ref {i2sums}) converge and a term $\sim r_{ij}^{d/2}$
comes from the asymptotic behavior of $I_2^{(d)}$.
Evaluating this, we can write the result in the form
\begin{eqnarray}
&& \frac{2 \lambda_{ij} I_2^{(d)} }{\Gamma\left(1-\frac{d}{2}\right)} =
   - \frac{\sqrt{\pi}\Gamma\left(\frac{d}{2}\right)}
 {\Gamma\left(\frac{d-1}{2}\right)}
 r_{ij}^{\frac{d-2}{2}}
 \left[ \frac{\partial_i  \lambda_{ij}}
        {\sqrt{1-\frac{m_j^2}{r_{ij} }}}
       +\frac{\partial_j \lambda_{ij}}
        {\sqrt{1-\frac{m_i^2}{r_{ij} }}} \right]
 \nonumber \\
&& \nonumber \\
&&~~ + \frac{\partial_i \lambda_{ij}}{(m_j^2)^{1-\frac{d}{2}}}
    \sum_{r=0}^{\infty} \frac{\left( \frac{d-1}{2}
   \right)_r}{\left(\frac{d}{2}\right)_r}
   \left( \frac{m_j^2}{r_{ij}}\right)^r 
   +\frac{\partial_j \lambda_{ij}}{(m_i^2)^{1-\frac{d}{2}}}
   \sum_{r=0}^{\infty} \frac{ \left(\frac{d-1}{2}\right)_r}
   {\left(\frac{d}{2}\right)_r}\left(\frac{m_i^2}{r_{ij}}
   \right)^r. 
\label{I2sums}
\end{eqnarray}
For $-(m^2_j-m^2_i) < p_{ij}^2 < 0$ we have again 
$r_{ij} > m_i^2,m_j^2$, but $\overline{x}_1 < 0 $ ,
 so that there
is no extremum inside the integration region and as a consequence
no contribution $\sim  r_{ij}^{d/2}$ comes from $I_2^{d}$ and the
boundary term is zero. In fact in this case the two contributions 
in the boundary term of (\ref {I2sums}) cancel and the formula is 
still valid. 

   Investigating as next $p_{ij}^2 > 0$ we can make the general
observation that $r_{ij} < m_i^2,m_j^2$, but $r_{ij} < 0$ is possible .
Thus, continuing
further to $0 < p_{ij}^2 < (m_j-m_i)^2$ we have $- \infty < r_{ij} < 0$,
but independent of its absolute value this does not cause problems
since even if the series in (\ref {I2sums}) do not converge, for
negative arguments the hypergeometric functions remain in their
domain of analyticity (also the boundary term vanishes) and (\ref{I2sums}) 
is valid if written in the form
\begin{eqnarray}
&& \frac{2 \lambda_{ij}~ I_2^{(d)} }{\Gamma\left(1-\frac{d}{2}\right)} =
   -\frac{\sqrt{\pi}\Gamma\left(\frac{d}{2}\right)}
 {\Gamma\left(\frac{d-1}{2}\right)}
 r_{ij}^{\frac{d-2}{2}}
 \left[ \frac{\partial_i  \lambda_{ij}}
        {\sqrt{1-\frac{m_j^2}{r_{ij}}}}
       +\frac{\partial_j \lambda_{ij}}
        {\sqrt{1-\frac{m_i^2}{r_{ij}}}} \right]
\nonumber \\
&& \nonumber \\
&&~~ + 
  \frac{\partial_i \lambda_{ij}}{(m_j^2)^{1-\frac{d}{2}}}
  \Fh21\Fdz{1,\frac{d-1}{2}}{\frac{d}{2}}
    + \frac{\partial_j \lambda_{ij}}
  {(m_i^2)^{1-\frac{d}{2}}}
  \Fh21\Fde{1,\frac{d-1}{2}}{\frac{d}{2}}.
\label{regionI}
\end{eqnarray}
By means of \cite{Erdeli}
\begin{eqnarray}
&&\Fh21\Ffz{1,\frac{d-1}{2}}{\frac{d}{2}} =
(2-d) \Fh21\Fzz{1,\frac{d-1}{2}}{\frac{3}{2}}
+\frac{\Gamma\left(\frac{d}{2}\right) \Gamma\left(\frac12\right)}
{\Gamma\left(\frac{d-1}{2}\right)}
\frac{z^{\frac{2-d}{2}}}{\sqrt{1-z}}, \\
&&~~~~~~~~~~~~~~~~~~~~~~~~~~~~~~~~~~~~~~~~~~~~~~~~~~~~~~~~~~~~~
\left| {\rm arg}(1-z)\right| < \pi
\nonumber 
\end{eqnarray}
a compact expression is obtained from (\ref{regionI}):
\begin{equation}
\frac{ \lambda_{ij}~ I_2^{(d)} }{\Gamma\left(2-\frac{d}{2}\right)} =
  \frac{\partial_i \lambda_{ij}}{(m_j^2)^{1-\frac{d}{2}}}
  \Fh21\Fdzz{1,\frac{d-1}{2}}{\frac{3}{2}}
+ \frac{\partial_j \lambda_{ij}}
  {(m_i^2)^{1-\frac{d}{2}}}
  \Fh21\Fdee{1,\frac{d-1}{2}}{\frac{3}{2}},
\label{MoRshort}
\end{equation}
which applies as before for $r_{ij} > m_i^2,m_j^2$, i.e. the
series for the hypergeometric functions do still converge, but beyond
that it is a valid representation also when we proceed to 
$(m_j-m_i)^2 < p_{ij}^2 < (m_j+m_i)^2$ since $0 < r_{ij} < m_i^2,m_j^2 $
in this case,
i.e. the argument of the hypergeometric function gets negative 
and therefore is in the domain of analyticity. 

Above the threshold, i.e. $(m_j+m_i)^2 < p_{ij}^2$, we have again 
$r_{ij} < 0$. Due to (\ref{rij}) for $p_{ij}^2 \rightarrow +\infty$, 
$r_{ij}$ behaves as
$r_{ij} \rightarrow -\infty$. Since in this case $0 < \overline{x}_1 < 1$
and $h_2$ has a minimum,
apparently for some value of $ p_{ij}^2$ we have $ |r_{ij}| > m_i^2,m_j^2 $
and the modulus of the argument of the hypergeometric function
in (\ref{regionI}) is $< 1$. Therefore, according to our approach,
the boundary term given in (\ref{regionI}) is also obtained in this
kinematical domain.

A very useful transformation-formula, applied to
(\ref{regionI}), is 
%%%%%%%%%%%%%%%%%%%%%%%%%%%%%%%%%%%%%%%%%%%%%%%%%%%%%
\begin{equation}
  \Fh21\Ffz{1,\frac{d-1}{2}}{\frac{d}{2}}=
  \frac{1}{1-z}\Fh21\Ffe{1,\frac12}{\frac{d}{2}},
\label{Feps}
\end{equation}
%%%%%%%%%%%%%%%%%%%%%%%%%%%%%%%%%%%%%%%%%%%%%%%%%%%%%
i.e. for all $r_{ij} < 0$ the arguments of the hypergeometric functions
can be transformed into values of modulus less than 1 such that the 
series converge.

To conclude, we have found for all $- \infty < p_{ij}^2 < + \infty$
a valid representation for $I_2^{(d)}$ and except for 
$0 < r_{ij} < m_i^2,m_j^2 $ even convergent series. If we wish
to obtain a convergent series in the latter case as well, we have to 
proceed as described in Sect. 3, 
(\ref{solution1a}), i.e. to perform the summation with $l < 0$. 
Without repeating the calculation in detail,
we here just give the result:
\begin{eqnarray}
&&\frac{g_{ij} I_2^{(d)}} {\Gamma \left( 2- \frac{d}{2} \right)} =
- \frac{ \Gamma \left( \frac32 \right) \Gamma \left( \frac{3-d}{2} \right)}
{\Gamma \left( 2- \frac{d}{2} \right)}
\left[ \frac{ \partial_i \lambda_{ij}}
                                    {\sqrt{m_j^2-r_{ij}}}
                           + \frac{ \partial_j \lambda_{ij}}
                                    {\sqrt{m_i^2-r_{ij}}}
                       \right]r_{ij}^{\frac{d-3}{2}}  \nonumber \\
&&-\frac{ \partial_i \lambda_{ij}} {(d-3) (m_j^2)^{\frac{4-d}{2}}}
 \Fh21\Frmj{1,2-\frac{d}{2}}{\frac{5-d}{2}}
  -\frac{ \partial_j \lambda_{ij}}{(d-3) (m_i^2)^{\frac{4-d}{2}}}
 \Fh21\Frmi{1,2-\frac{d}{2}}{\frac{5-d}{2}}. 
\label{rijm}
\end{eqnarray}
Using
\begin{eqnarray}
&& \Fh21\Ffz{1,2-\frac{d}{2}}{\frac{5-d}{2}}=
(d-3) 
\Fh21\Fzz{1,2-\frac{d}{2}}{\frac{3}{2}}
+\frac{\Gamma\left(\frac{5-d}{2}\right) \Gamma\left(\frac12\right)}
{\Gamma\left(\frac{4-d}{2}\right)}
\frac{z^{\frac{d-3}{2}}}{\sqrt{1-z}}
\nonumber \\
&&~~~~~~~~~~~~~~~~~~~~~~~~~~~~~~~~~~~~~~~~~~~~~~~~~~~~~~~~~~
\left| {\rm arg}(1-z)\right| <\pi . 
\label{Geps}
\end{eqnarray}
yields from (\ref{rijm}) the already known result \cite{BDS}:
\begin{equation}
\frac{g_{ij}~I_2^{(d)}}
  { \Gamma\left(2-\frac{d}{2}\right)} = -
 \frac{\partial_i \lambda_{ij}}
     { (m_j^2)^{\frac{4-d}{2}}}
     \Fh21\Fdj{1,\frac{4-d}{2}}{\frac{3}{2}}
  -\frac{\partial_j \lambda_{ij}}
      { (m_i^2)^{\frac{4-d}{2}}} 
     \Fh21\Fdi{1,\frac{4-d}{2}}{\frac{3}{2}}.
\label{RoMshort}
\end{equation}     

Having thus obtained different representations for $I_2^{(d)}$
by different ways of solving the difference equation (see Sect. 3), 
we can also show that they agree: by means of
%%%%%%%%%%%%%%%%%%%%%%%%%%%%%%%%%%%%%%%%%%%%%%%%%%%%%
\begin{equation}
\Fh21\Fzz{1,\frac{d-1}{2}}{\frac{3}{2}}=
z^{-1}
\Fh21\Foz{1,\frac{4-d}{2}}{\frac{3}{2}}
\end{equation}
%%%%%%%%%%%%%%%%%%%%%%%%%%%%%%%%%%%%%%%%%%%%%%%%%%%%%
(\ref{RoMshort}) can be obtained from (\ref{MoRshort}).\\

For future applications we also give the 
$\varepsilon$ expansion of $_2F_1$ occurring in (\ref{Feps})
(see \cite{Fleischer:1998dw} formula A.3 in the Appendix):
\begin{eqnarray}
&&     \Fh21\Ffx{1,\frac{1}{2}}{\frac{d}{2}}=
     \frac{(1-\varepsilon) (1-y^2)}{(1-2\varepsilon ) y}\left[ 
\frac{y}{1-y}
 +\ln(1-y) \varepsilon
 -\left( {\rm Li}_2(y)+ \ln^2(1-y) \right) \varepsilon^2
\right.
\nonumber \\
&&\left.
+\left( 2 S_{12}(y)+2 \ln(1-y) {\rm Li}_2(y)- {\rm Li}_3(y)
 +\frac23 \ln^3(1-y) \right)\varepsilon^3
 \right] + O(\varepsilon^4),
\end{eqnarray}
where
\begin{equation}
y=\frac{1-\sqrt{1-x}}{1+\sqrt{1-x}}.
\end{equation}
The $\varepsilon$ expansion of $_2F_1$ in (\ref{Geps})
can be obtained from
\begin{eqnarray}
&& \Fh21\Fzz{1,2-\frac{d}{2}}{\frac{3}{2}}=
\frac{1}{z (2-d)}
\Fh21\Fiz{1,\frac{d-1}{2}}{\frac{d}{2}}
+\frac{\Gamma\left(\frac{d-2}{2}\right) \Gamma\left(\frac32\right)}
{\Gamma\left(\frac{d-1}{2}\right)}
\frac{z^{\frac{d-3}{2}}}{\sqrt{{z-1}}}.
\end{eqnarray}
More information on the $\varepsilon$ expansion of hypergeometric
functions is given in \cite{Davydychev:2000na}.
%%%%%%%%%%%%%%%%%%%%%%%%%%%%%%%%%%%%%%%%%%%%%%%%%%%%%%%%%%%%%%
%%%%%%%%%%%%%%%%%%%%%%%%%%%%%%%%%%%%%%%%%%%%%%%%%%%%%%%%%%%%%%
\section{ 3-point function}
%%%%%%%%%%%%%%%%%%%%%%%%%%%%%%%%%%%%%%%%%%%%%%%%%%%%%%%%%%%%%%
%%%%%%%%%%%%%%%%%%%%%%%%%%%%%%%%%%%%%%%%%%%%%%%%%%%%%%%%%%%%%%

   In complete analogy to (\ref {i2sums}) we proceed for the
3-point function. In this case we have to sum according to
(\ref {npoint}) 2-point functions, which in Sect. 4 were
represented in terms of hypergeometric functions $_2F_1$.
In order to simplify the summation, we eliminate $d$ in the first
argument of $_2F_1$ by means of
\begin{equation}
  \Fh21\Ffz{a,b}{c}=(1-z)^{-a}\Fh21\Ffe{a,c-b}{c}.
\label{Simpl}
\end{equation}
Dissolving the result into double series, it can be written as 
Appell hypergeometric functions $F_3( (d-2)/2,1,1,1/2,d/2;x,y)$ 
defined as
\begin{equation}
    F_3(\alpha, \alpha', \beta,\beta',\gamma;x,y)=
   \sum_{m,n=0}^{\infty}\frac{(\alpha)_{m}(\alpha')_{n}
   (\beta)_{m}(\beta')_{n}}{(\gamma)_{m+n}~ m! ~n!}x^m y^n,
\end{equation}
which in turn can be reduced to $F_1$ by means of \cite{Erdeli}
%&&&&&&&&&&&&&&&&&&&&&&&&&&&&&&&&&&&&&&&&&&&&&&&&&&&&&&&&&&&&&
\begin{equation}
  F_3(\alpha, \alpha', \beta,\beta',\alpha+\alpha';x,y)=
  (1-y)^{-\beta'} F_1 \left(\alpha,\beta,\beta',\alpha+\alpha';
  x,\frac{y}{y-1}\right),
\end{equation}
$F_1$ being defined as
\begin{equation}
  F_1(\alpha,\beta,\beta',\gamma;x,y)=\sum_{m,n=0}^{\infty}
  \frac{(\alpha)_{m+n}(\beta)_{m}(\beta')_{n}}
  {(\gamma)_{m+n}~ m! ~n!}x^m y^n.
\end{equation}
%&&&&&&&&&&&&&&&&&&&&&&&&&&&&&&&&&&&&&&&&&&&&&&&&&&&&&&&&&&&&&
In the next step the boundary term in (\ref {npoint}) has to  be
determined by means of the asymptotic method described in Sect. 3.1~.
The polynomial $h_3$ (see (\ref {RePlusIm}) for $n=3$) needed to
determine a possible extremum `inside' the integration region, 
is given by
\begin{eqnarray}
&&h_3=
  -x_1 x_2 (1-x_1) p^2_{ik}-x_1^2x_2 (1-x_2)  p^2_{ij}
  -x_1 (1-x_1)(1-x_2)  p^2_{jk}
  \nonumber \\
&&~~  
  +x_1 x_2 m_i^2+x_1 (1-x_2) m_j^2+(1-x_1) m_k^2.
\label{h3}
\end{eqnarray}
It has an extremum at
\begin{equation}
\overline{x}_1 = 
 -\frac{(\partial_i \lambda_{ijk}+\partial_j \lambda_{ijk} )}{G_2},~~~
\overline{x}_2=\frac{\partial_i \lambda_{ijk}}{\partial_i \lambda_{ijk}
+ \partial_j \lambda_{ijk}},
\end{equation}
provided the determinant
\begin{equation}
D = \frac{\partial^2 h_3}{\partial x_1 \partial x_1}
\frac{\partial^2 h_3}{\partial x_2 \partial x_2} -
\left( \frac{\partial^2 h_3}{\partial x_1 \partial x_2}
\right)^2 = 
-\frac{(\partial_i \lambda_{ijk}+ \partial_j \lambda_{ijk})^2}{2G_2}>0.
\label{Determi}
\end{equation}
Further, the following second derivatives determine the type of the extremum:
\begin{equation}
\frac{\partial^2 h_3}{\partial x_1 \partial x_1}
= \frac{2 G_2^2 (m_k^2-r_{ijk} )}
{(\partial_i \lambda_{ijk}+ \partial_j \lambda_{ijk})^2},
\label{x1x1}
\end{equation}
\begin{equation}
\frac{\partial^2 h_3}{\partial x_2 \partial x_2}
=\frac{2 p_{ij}^2 (\partial_i \lambda_{ijk} 
+ \partial_j \lambda_{ijk})^2}{G_2^2}.
\label{x2x2}
\end{equation}
If both are $ > 0,~ h_3$  has a local minimum, if both are $ < 0,~  h_3$
has a local maximum. Thus we see that a maximum will be achieved if
\begin{equation}
G_2<0, ~~r_{ijk} > m_i^2,m_j^2,m_k^2, ~~p_{ij}^2<0.
\label{achiev}
\end{equation}
Here the explicit appearance of $m_k$ in (\ref {x1x1}) 
and of $p_{ij}^2$ in (\ref {x2x2}) are due to the particular
choice of (\ref {h3}), i.e. the choice of the Feynman parameters. 
In general no preference of any choice of parameters can occur, 
hence the above condition for $r_{ijk}$ and $p_{ij}^2$.

The result for $I_3^{(d)}$ finally reads \cite{Tarasov:2000sf}
%
%%%%%%%%%%%%%%    main result for I_3^{(d)}
%
\begin{equation}
  \frac{\lambda_{ijk}}{\Gamma \left( 2-\frac{d}{2} \right)}
   I_3^{(d)}=b_3
  +\theta_{ijk}~\partial_k \lambda_{ijk}
  +\theta_{kij}~\partial_j \lambda_{ijk}
  +\theta_{jki}~\partial_i \lambda_{ijk},
\label{I3main}
\end{equation} 
%&&&&&&&&&&&&&&&&&&&&&&&&&&&&&&&&&&&&&&&&&&&&&&&&&&&&&&&&&&&
where 
\begin{equation}
b_3=2^{\frac32} \pi~\sqrt{-g_{ijk}}~ r_{ijk}^{\frac{d-2}{2}},
\label{b3}
\end{equation}
(see (\ref{npointEucl}) )
provided an extremum of $h_3$ ($G_2 < 0$) occurs inside the 
integration region of the Feynman parameters. Otherwise $b_3=0$. 
For $\lambda_{ij} \neq 0$ we have
\begin{eqnarray}
\lambda_{ij}~ \theta_{ijk}&=&-
 \left[ \frac{\partial_i  \lambda_{ij}}
        {\sqrt{1-\frac{m_j^2}{r_{ij}}}}
       +\frac{\partial_j \lambda_{ij}}
        {\sqrt{1-\frac{m_i^2}{r_{ij}}}} \right]
       r_{ij}^{\frac{d-2}{2}}~
  \frac{\sqrt{\pi}~ \Gamma\left(\frac{d-2}{2}\right)}
  {4 \Gamma\left(\frac{d-1}{2}\right)}
 \Fh21\Fdf{1,\frac{d-2}{2}}{\frac{d-1}{2}} \nonumber \\
&& \nonumber \\
&+&\frac{(m_i^2)^{\frac{d-2}{2}}}
  {2(d-2)}
  \frac{\partial_j\lambda_{ij}}{ \sqrt{1-\frac{m_i^2}{r_{ij}}}}
  F_1\left(\frac{d-2}{2},1,\frac12,
  \frac{d}{2}; \frac{m_i^2 }{r_{ijk}},
  \frac{m_i^2}{r_{ij}}\right) 
\nonumber \\
&& \nonumber \\
&+&\frac{(m_j^2)^{\frac{d-2}{2}}}
  {2(d-2)}
  \frac{\partial_i\lambda_{ij}}{ \sqrt{1-\frac{m_j^2}{r_{ij}}}}
  F_1\left(\frac{d-2}{2},1,\frac12,
  \frac{d}{2}; \frac{m_j^2 }{r_{ijk}},
  \frac{m_j^2}{r_{ij}}\right).
\label{TetaForI3}
\end{eqnarray}
To our knowledge there exists no simpler hypergeometric 
representation of the $3-$point function for $d$ dimensions
in the literature. For another approach see e.g. \cite{Davydychev:1997wa}.
The function $F_1$ admits a simple integral representation:
\begin{equation}
F_1(\alpha,\beta,\beta',\gamma; x,y)=
 \frac{\Gamma(\gamma) }{\Gamma(\alpha)\Gamma(\gamma-\alpha)}
\int_0^1 \frac{u^{\alpha-1} (1-u)^{\gamma-\alpha-1}}
{(1-xu)^{\beta}(1-yu)^{\beta'}}~du,
\end{equation}
which in our case is
\begin{equation}
F_1 \left( \frac{d-2}{2},1,\frac12,\frac{d}{2}; x,y \right)=
 \frac{d-2}{2}
\int_0^1 \frac{u^{\frac{d-4}{2}}} 
{(1-xu) \sqrt{1-yu} }du.
\label{intrepF1}
\end{equation}
For further information on $F_1$, we mention for its analytic
continuation \cite{olsson} and concerning methods for its numerical
evaluation \cite{numericalF1}.
%%%%%%%%%%%%%%%%%%%%%%%%%%%%%%%%%%%%%%%%%%%%%%%%%%%%%%%%%%%%%%%%
For $\lambda_{ij}$ arbitrary, using the transformation formula 
\begin{eqnarray}
&&F_1 \left(\frac{d-2}{2},1,\frac12,\frac{d}{2};x,y\right)
=\frac{\sqrt{\pi}\Gamma\left( \frac{d}{2}\right) }
{\Gamma\left( \frac{d-1}{2}\right)} y^{\frac{2-d}{2} }
 \Fh21\Fxy{1,\frac{d-2}{2}}{\frac{d-1}{2}}
\nonumber \\
&&~~+\frac{(2-d) ~\sqrt{1-y}}{(1-x) y}
 F_1 \left(1,2-\frac{d}{2},1,\frac32;1-\frac{1}{y},
 \frac{x(1-y)}{y(1-x)}\right)
\label{A1}
\end{eqnarray}
the following compact form for $\theta_{ijk}$ is obtained:
\begin{eqnarray}
g_{ij}~ \theta_{ijk}=
&+&\partial_j\lambda_{ij}\frac{(m_i^2)^{\frac{d-4}{2}}  }
  {2 \left(1-\frac{m_i^2}{r_{ijk}}\right)  }~
  F_1\left(1,2-\frac{d}{2},1,\frac32
  ; 1-\frac{r_{ij}}{m_i^2 },
  \frac{r_{ij}-m_i^2}{r_{ijk}-m_i^2}\right) 
\nonumber \\
&& \nonumber \\
&+&\partial_i\lambda_{ij} \frac{(m_j^2)^{\frac{d-4}{2}}  }
  {2 \left(1-\frac{m_j^2}{r_{ijk}}\right) }~
  F_1\left(1,2-\frac{d}{2},1,\frac32
  ; 1-\frac{r_{ij}}{m_j^2 },
  \frac{r_{ij}-m_j^2}{r_{ijk}-m_j^2}\right).
\label{I3comp}
\end{eqnarray}
%----------------------------------------------------------------
where again $F_1$ has a simple integral representation
\begin{equation}
F_1\left(1,2-\frac{d}{2},1,\frac32
  ; x,y\right) = \frac12 \int_0^1 \frac{du}{\sqrt{1-u}} 
  \frac{(1-xu)^{\frac{d-4}{2}}}{(1-yu)}.
\end{equation}  
To find $I_3^{(d)}$ up to first order in $\varepsilon=(4-d)/2$, 
three terms in the expansion of this function are needed
\begin{eqnarray}
&&F_1\left(1,2-\frac{d}{2},1,\frac32
  ; x,y\right) =-\frac{2B}{1-B^2}\left\{ \ln B \right.
\nonumber \\
&& ~~ 
 +\varepsilon \left[ {\rm Li}_2(1-AB)+{\rm Li}_2\left(1-\frac{B}{A}\right)
 -2{\rm Li}_2(1-B)+\frac12 \ln^2A\right]
\nonumber \\
&& +\varepsilon^2 \left[
      {\rm Li}_3\left( \frac{A(1-AB)}{A-B}\right)
     -{\rm Li}_3\left( \frac{A(A-B)}{1-AB}\right)
    +2{\rm Li}_3\left( \frac{A(1-B)}{1-AB}\right)
\right.
\nonumber \\
&&~~~~~~
    -2{\rm Li}_3\left( \frac{A(1-B)}{A-B}\right)
    +2{\rm Li}_3\left( \frac{1-B}{A-B}\right)
    -2{\rm Li}_3\left( \frac{1-B}{1-AB}\right) \nonumber \\
&&+ 2\left[{\rm Li}_2\left( \frac{A(A-B)}{1-AB}\right)
          -{\rm Li}_2\left( \frac{A(1-B)}{1-AB}\right)
          +{\rm Li}_2\left( \frac{1-B}{A-B}\right)
          -{\rm Li}_2(-A)
    \right] \ln(A)
\nonumber \\
&&~~ 
+\left[\frac12\ln^2(A)-\zeta(2)\right]\ln\left(\frac{B-A}{1-AB}\right)
-\frac16 \ln^3\left(\frac{B-A}{1-AB}\right)
\nonumber \\
&&~~\left.\left.
+\frac12\ln (A)\ln^2 \left(\frac{B-A}{1-AB}\right)
\right] +O(\varepsilon^3)
\right\}.
\end{eqnarray}
where
\begin{equation}
   A=\frac{\sqrt{1-\frac{1}{x}}-1}{\sqrt{1-\frac{1}{x}}+1},
~~~B=\frac{\sqrt{1-\frac{1}{y}}-1}{\sqrt{1-\frac{1}{y}}+1}.
\end{equation}
This expansion will be of interest for an evaluation of one-loop
counterterms inserted into vertex functions, required in two-loop
calculations.

   The general result(\ref {I3main}) for the 3-point function
with arbitrary masses and external momenta is finally a bit lengthy.
To demonstrate the usefulness of our result, we choose as a particular
case the on-shell 3-point function occurring in Bhabha scattering:
$J_G^{(d)}$ with $m_1^2=m_2^2=0$, $m_3^2=m^2$ and $p_1^2=p_2^2=m^2$. 
For $0 < p_{12}^2 < 4 m^2$ we have $G_2 < 0$, i.e. an extremum (minimum)
exists, but $\overline{x}_1 = \frac{4 m^2}{4 m^2- p_{12}^2 }> 1$,
such that it is not `inside'. Therefore $b_3=0$ in this case.

   Evaluating the three remaining terms in (\ref {I3main}), we put
$i,j,k = 1,2,3$, i.e. we have to calculate $\theta_{123}, \theta_{312}$
and $\theta_{231}$. With $\lambda_{12}=-p_{12}^2$ 
and $\lambda_{31}=\lambda_{23}=0$ we see that for the two latter 
cases we have to use (\ref {I3comp}). In the
first case only the hypergeometric function $\Fh21$ remains:
\begin{equation}
\theta_{123}=-
  \frac{\sqrt{\pi}~ \Gamma\left(\frac{d-2}{2}\right)}
  {2^{d-2} \Gamma\left(\frac{d-1}{2}\right)}
  (-p_{12}^2)^{\frac{d-4}{2}}
 \Fh21\Fdg{1,\frac{d-2}{2}}{\frac{d-1}{2}}
\end{equation}
For $\theta_{312}$ and $\theta_{231}$, using (\ref {I3comp}), only
one of the $F_1$ functions remains, respectively, since 
$\partial_3\lambda_{31}=0$ and $\partial_3\lambda_{23}=0$. 
Moreover the two last contributions in (\ref {I3main}) are the same. 
Evaluating thus $\theta_{312}$ we obtain:
\begin{equation}
\theta_{312}=-\frac{1}{2} (m^2)^{\frac{d-4}{2}} \frac{p_{12}^2}{4 m^2}
F_1\left(1,1,2-\frac{d}{2},\frac{3}{2};1-\frac{p_{12}^2}{4 m^2},1 \right).
\end{equation}
With $\lambda_{123}=-2 m^2 (p_{12}^2)^2$, 
$\partial_3\lambda_{123}=-2 p_{12}^2$, 
$\partial_1\lambda_{123}=\partial_2\lambda_{123}=4 m^2 p_{12}^2$,
and
\begin{equation}
F_1(a,b,b^{'},c,w,1)= 
\frac{\Gamma(c) \Gamma(c-a-b^{'})}{\Gamma(c-a) \Gamma(c-b^{'})} \nonumber
\Fh21\Ffw{a,b}{c-b^{'}}
\end{equation}
the final result is
\begin{equation}
\frac{J_G} {\Gamma \left(2-\frac{d}{2}\right)}= 
  \frac{(m^2)^{\frac{d}{2}-3}}
          {2(d-3)} 
       \Fh21\Fpm{1,1}{\frac{d-1}{2}}
 - \frac{ \sqrt{\pi} \Gamma \left(\frac{d-2}{2}\right)
         (-p^2)^{\frac{d-4}{2}}}
         {2^{d-2} \Gamma \left(\frac{d-1}{2}\right) m^2}
        \Fh21\Fpm{1,\frac{d-2}{2}}{\frac{d-1}{2}}.
\label{J_G}
\end{equation}
The result for $J_G$ at $d=4$ was given in \cite{Davydychev:2000rt}.
%%%%%%%%%%%%%%%%%%%%%%%%%%%%%%%%%%%%%%%%%%%%%%%%%%%%%%%%%%%%%
%%%%%%%%%%%%%%%%%%%%%%%%%%%%%%%%%%%%%%%%%%%%%%%%%%%%%%%%%%%%%
\section{ 4-point function}
%%%%%%%%%%%%%%%%%%%%%%%%%%%%%%%%%%%%%%%%%%%%%%%%%%%%%%%%%%%%%
%%%%%%%%%%%%%%%%%%%%%%%%%%%%%%%%%%%%%%%%%%%%%%%%%%%%%%%%%%%%%
   The calculation of the $4-$point function follows the same scheme
as has been described in detail in the foregoing sections. 
Concerning the boundary term, it is in principle not difficult to 
find a relative extremum in a cube, but to find a proper formulation 
of all conditions (as has been done in the case of the $3-$point 
function, see (\ref{Determi})~-~(\ref{achiev}))
in terms of `standard' expressions like $G_3$ etc. turns out to be tedious.
However, due to the discussion after (\ref{xbarli}), one excludes 
an extremum `inside' if not all derivatives $\partial_k \Delta_n$ have
the same sign.  In the case of the $4-$point Bhabha diagrams, nevertheless, 
we demonstrate an alternative approach to determine the boundary term.

   We label the lines of the box integral in natural manner as 
$i,j,k,l$ . As one particular case we select an $I_3^{(d)}$ 
obtained  from $I_4^{(d)}$ by shrinking line $l$ and take into 
account at first only the first $F_1$ term in (\ref{TetaForI3}):
\begin{equation}
I^{(d)}_{4,F_1}=
-\left(\frac{\partial_l \lambda_{ijkl}}{2 \lambda_{ijkl}}\right)
\sum_{r=0}^{\infty}\left(\frac{d-3}{2}\right)_r
\left( \frac{g_{ijkl}}{\lambda_{ijkl}} \right)^r I_{3,F_1}^{(d+2r)}.
\end{equation}
In analogy to (\ref{Simpl}) for the $3-$point function, 
to simplify the summation, we get rid of $d$ in the
first argument of $F_1$ by means of the transformation formula
\cite{Erdeli}
\begin{equation}
F_1(\alpha,\beta, \beta',\gamma;x,y)
 =(1-x)^{-\beta}(1-y)^{-\beta'}F_1\left(\gamma-\alpha, 
\beta,\beta',\gamma; \frac{x}{x-1},\frac{y}{y-1}\right).
\end{equation}
After this simplification we arrive at
\begin{eqnarray}
&&I^{(d)}_{4,F_1}= \frac18 \left(\frac{\partial_l \lambda_{ijkl}}
                                       {\lambda_{ijkl}}\right)
\left(\frac{\partial_k \lambda_{ijk}}
              {\lambda_{ijk}}\right)
\left(\frac{\partial_j \lambda_{ij}}
            {\lambda_{ij}}\right) 
\frac{(m_i^2)^{\frac{d-2}{2}} r_{ijk}}{r_{ijk}-m_i^2}
       \frac{r_{ij}}{r_{ij}-m_i^2} 
\Gamma \left(1-\frac{d}{2}\right)
\nonumber \\
&&~\sum_{r,m,n=0}^{\infty}
\frac{ \left( 
\frac{d-3}{2}\right)_r (1)_{m+n} (1)_r (1)_m \left( \frac12 \right)_n }
{ \left(\frac{d}{2}\right)_{r+m+n}~m! n! r!}
\left( \frac{m_i^2}{r_{ijkl}}\right)^r
\left( \frac{m_i^2}{m_i^2-r_{ijk}}\right)^m
\left( \frac{m_i^2}{m_i^2-r_{ij}}\right)^n
\nonumber \\
&&~~ =  \frac18 \left(\frac{\partial_l \lambda_{ijkl}}
                                       {\lambda_{ijkl}}\right)
\left(\frac{\partial_k \lambda_{ijk}}
              {\lambda_{ijk}}\right)
\left(\frac{\partial_j \lambda_{ij}}
            {\lambda_{ij}}\right) 
\frac{(m_i^2)^{\frac{d-2}{2}} r_{ijk}}{r_{ijk}-m_i^2}
        \frac{r_{ij}}{r_{ij}-m_i^2} 
\Gamma \left(1-\frac{d}{2}\right)
\nonumber \\
&&~~~\times F_S\left( \frac{d-3}{2},1,1; 1,1,\frac12;
\frac{d}{2},\frac{d}{2},\frac{d}{2};
\frac{m_i^2}{r_{ijkl}},
\frac{m_i^2}{m_i^2-r_{ijk}},
\frac{m_i^2}{m_i^2-r_{ij}} \right),
\end{eqnarray}
where the Lauricella-Saran function $F_S$ is given in terms
of a triple hypergeometric series \cite{Lauricella},
\cite{Saran54}:
\begin{eqnarray}
&&F_S(\alpha_1,\alpha_2,\alpha_2,
       \beta_1, \beta_2, \beta_3,
      \gamma_1,\gamma_1,\gamma_1; x,y,z) \nonumber \\
&&~~=\sum_{r,m,n=0}^{\infty}
 \frac{(\alpha_1)_r (\alpha_2)_{m+n} (\beta_1)_{r}
 (\beta_2)_m (\beta_3)_n }{(\gamma_1)_{r+m+n}~ r! m! n!}
 x^r y^m z^n.
\end{eqnarray}
In our case the following integral representation of
$F_S$ is useful \cite{Saran55}
\begin{eqnarray}
&&\frac{\Gamma(\alpha_1)\Gamma(\gamma_1-\alpha_1)}
  {\Gamma(\gamma_1)}F_S(\alpha_1,\alpha_2,\alpha_2,
  \beta_1,\beta_2,\beta_3; \gamma_1,\gamma_1,\gamma_1;x,y,z)
  = \nonumber\\
&& \nonumber \\
&&~~\int_0^1 \frac{t^{\gamma_1-\alpha_1-1} (1-t)^{\alpha_1-1}}
  {(1-x+tx)^{\beta_1}}F_1(\alpha_2,\beta_2,\beta_3;
  \gamma_1-\alpha_1;ty,tz) dt,
\label{FSintF1}
\end{eqnarray}
which, specified for the above parameters, yields
\begin{eqnarray}
&&F_S\left( \frac{d-3}{2},1,1; 1,1,\frac12;
\frac{d}{2},\frac{d}{2},\frac{d}{2};
x,y,z \right)=\frac{\Gamma \left( \frac{d}{2}\right)}
{\Gamma \left( \frac{d-3}{2}\right)\Gamma \left( \frac{3}{2}\right)}
\nonumber \\
&&~\times \int_0^1 \frac{dt~\sqrt{t}(1-t)^{\frac{d-5}{2}}}
{(1-x+tx)} F_1\left(1,1,\frac12,\frac32; ty,tz\right).
\end{eqnarray}
By means of 
\begin{equation}
F_1\left(a,b,b',b+b'; w,z\right)=(1-z)^{-a} \Fh21\Fwz{a,b}{b+b'}
\end{equation}
$F_1$ in the integrand reduces to
\begin{equation}
F_1\left(1,1,\frac12,\frac32;ty,tz\right) =
\frac{1}{1-tz} \Fh21\Fwzz{1,1}{\frac32}
=\frac{\arcsin \sqrt{\frac{t(y-z)}{1-tz}}}{\sqrt{t(y-z)(1-ty)}}
\end{equation}
and finally we obtain
\begin{eqnarray}
&&F_S\left( \frac{d-3}{2},1,1; 1,1,\frac12;
\frac{d}{2},\frac{d}{2},\frac{d}{2};
x,y,z \right) \nonumber \\
&&~~~=\frac{\Gamma \left(\frac{d}{2}\right)~(y-z)^{-\frac12}  }
{\Gamma\left(\frac{d-3}{2}\right) \Gamma\left(\frac32\right)}
 \int_0^1 \frac{\arcsin \sqrt{\frac{(y-z)t}{1-tz}}
 (1-t)^{\frac{d-5}{2}}}{(1-x+tx) \sqrt{1-ty}} ~dt .
\label{intrepFS}
\end{eqnarray}
Secondly the contribution to $I_4^{(d)}$ from $_2F_1$ in
(\ref{TetaForI3}) can be evaluated in the same manner as 
in Sect. 5, where $I_3^{(d)}$ was calculated from $I_2^{(d)}$. 
The result reads:
\begin{eqnarray}
&&I_{4,_2F_1}^{(d)} = 
\frac{\sqrt{\pi}}{8} \left(\frac{\partial_l \lambda_{ijkl}}
                                       {\lambda_{ijkl}}\right)
\left(\frac{\partial_k \lambda_{ijk}}
              {\lambda_{ijk}}\right)
\frac{(r_{ij})^{\frac{d-2}{2}}} {\lambda_{ij}
\sqrt{1-\frac{r_{ij}}{r_{ijk}}}    }
\frac{
\Gamma \left(\frac{4-d}{2}\right)
\Gamma \left(\frac{d-2}{2}\right)}
{\Gamma \left(\frac{d-1}{2}\right)}
\nonumber \\
&&~~ \times  \left[ \frac{\partial_i  \lambda_{ij}}
        {\sqrt{1-\frac{m_j^2}{r_{ij}}}}
       +\frac{\partial_j \lambda_{ij}}
        {\sqrt{1-\frac{m_i^2}{r_{ij}}}} \right]
F_1\left( \frac{d-3}{2},1,\frac12,\frac{d-1}{2};
\frac{r_{ij}}{r_{ijkl}},\frac{r_{ij}}{r_{ijk}} \right).
\end{eqnarray}
In complete analogy to (\ref{I3main}) we now write
the result for the $4$-point function as
\begin{eqnarray}
  \frac{\lambda_{ijkl}}{\Gamma\left( 2-\frac{d}{2}\right)}
  I_4^{(d)} &=& b_4
\nonumber \\
&& \nonumber \\
&+&\phi_{ijkl} ~\partial_l \lambda_{ijkl}
     +\phi_{lijk} ~\partial_k \lambda_{ijkl}
     +\phi_{klij} ~\partial_j \lambda_{ijkl}
     +\phi_{jkli} ~\partial_i \lambda_{ijkl},
\end{eqnarray}     
where
\begin{eqnarray}
&&\phi_{ijkl}=\left\{-\frac{\pi~\sqrt{2}}{\sqrt{-g_{ijk}}}
  ~r_{ijk}^{\frac{d-4}{2}}~
  \Fh21\Fbf{1,\frac{d-3}{2}}{\frac{d-2}{2}} \right. \nonumber \\
&& \nonumber \\
&& + \frac{1}{8}
\left(\frac{\partial_k \lambda_{ijk}}
              {\lambda_{ijk}}\right)
\frac{1}{\lambda_{ij}}
\left[  \frac{\sqrt{\pi} \Gamma\left(\frac{d-2}{2}\right)}
                        {\Gamma\left(\frac{d-1}{2}\right)}
\frac{r_{ij}^{\frac{d-2}{2}}}{\sqrt{1-\frac{r_{ij}}{r_{ijk}}}}
 \left( \frac{\partial_i  \lambda_{ij}}
        {\sqrt{1-\frac{m_j^2}{r_{ij}}}}
       +\frac{\partial_j \lambda_{ij}}
        {\sqrt{1-\frac{m_i^2}{r_{ij}}}} \right)
\right. 
\nonumber \\
&& ~~\times ~F_1\left( \frac{d-3}{2},1,\frac12,\frac{d-1}{2};
\frac{r_{ij}}{r_{ijkl}},\frac{r_{ij}}{r_{ijk}} \right)
\nonumber \\
&&~~-\frac{2 (m_i^2)^{\frac{d-2}{2}}}{d-2}(\partial_j \lambda_{ij}) 
\frac{r_{ijk}}{r_{ijk}-m_i^2} 
\frac{r_{ij }}{r_{ij }-m_i^2}
~F_S \left( \frac{m_i^2}{r_{ijkl}},
\frac{m_i^2}{m_i^2-r_{ijk}},
\frac{m_i^2}{m_i^2-r_{ij}} \right)
\nonumber \\
&&~~\left.\left. 
-\frac{2 (m_j^2)^{\frac{d-2}{2}}}{d-2}(\partial_i \lambda_{ij}) 
\frac{r_{ijk}}{r_{ijk}-m_j^2} 
\frac{r_{ij }}{r_{ij }-m_j^2}
~F_S\left( \frac{m_j^2}{r_{ijkl}},
\frac{m_j^2}{m_j^2-r_{ijk}},
\frac{m_j^2}{m_j^2-r_{ij}} \right)
\right] \right\}
\nonumber\\
&& \nonumber \\
&&~~~~~~~~~~+ \left\{ i,j,k ~~~\rightarrow~~~{k,i,j} \right\}
          + \left\{ i,j,k ~~~\rightarrow~~~{j,k,i} \right\}
\label{Lafo4}
\end{eqnarray}
The first term on the r.h.s. of (\ref{Lafo4}) comes from $b_3$
of (\ref{I3main}), but is supposed to be $0$ if the $3-$point
function is evaluated in a kinematical domain, where no boundary
term occurs.
The boundary term $b_4$ we give for completeness:
\begin{equation}
b_4=8 {\pi}^{\frac32}~\frac{\Gamma\left(\frac{d}{2}\right)}
{(d-2) \Gamma\left(\frac{d-3}{2}\right)}
\sqrt{-g_{ijkl}}~ r_{ijkl}^{\frac{d-3}{2}}
\end{equation}
with the same reasoning as for (\ref{b3}).
So far this result demonstrates our general method also for the
4-point function.

To give an explicit and complete example 
we consider now the case with the following kinematical
variables
\begin{eqnarray}
&&p_1^2=m^2,~~~  p_2^2=s,~~~p_3^2=m^2,~~~p_1p_2=p_2p_3=\frac12 s,
~~~ p_1p_3=m^2-\frac12 t, 
\nonumber \\
&&~m_1^2=m_3^2=0,~~~m_2^2=m_4^2=m^2,
\label{Bhabkiva}
\end{eqnarray}
which corresponds to the (scalar) box diagram with two
photons in the s-channel, occurring in Bhabha scattering.
Apart from (\ref{J_G}) we also need the 3-point function
$J_F^{(d)}$ with $m_1^2=m_2^2=m^2$, $m_3^2=0$ and $p_1^2=p_2^2=m^2$,
which is
\begin{equation}
J_F^{(d)}=\frac{\Gamma\left(2-\frac{d}{2}\right)}{2 ~m^{6-d}}
\Fh21\Ffp{1,3-\frac{d}{2}}{\frac32}.
\end{equation}
With these two 3-point functions we can set up the difference
equation for the 4-point function with kinematics given in 
(\ref{Bhabkiva}), the solution of which finally reads (the indices of
$I_{1111}$ being the powers of the scalar propagators, see also
(\ref{nuieq1}))
\begin{eqnarray}
&&I_{1111}^{(d)} = \frac{b_4}{z^{d/2} \Gamma\left(\frac{d-3}{2}\right)}
\nonumber \\
&& - \frac{4 m^{d-4}}{t(s-4m^2)} \Gamma \left(2-\frac{d}{2}\right)
 F_2\left(\frac{d-3}{2},1,1,\frac32, \frac{d-2}{2};
 \frac{s}{s-4m^2}, -m^2z \right)
 \nonumber \\
&&+\frac{4 m^{d-4}}{(d-3)t(s-4m^2)}\Gamma\left(2- \frac{d}{2}\right)
F^{1;2;1}_{1;1;0} \left[^{\frac{d-3}{2}:~ \frac{d-3}{2},~1;~~~~ 1;}
_{\frac{d-1}{2}:~~~~~~ \frac{d-2}{2};~~-;}~~-m^2z,1-\frac{4m^2}{t}\right]
\nonumber \\
&& -\frac{\sqrt{\pi} (-t)^{\frac{d-4}{2}}}{2^{d-4}m  \sqrt{t}}
~\frac{\Gamma\left(\frac{d-2}{2}\right) \Gamma\left(2-\frac{d}{2}\right)}
{(s-4m^2) \Gamma\left( \frac{d-1}{2}\right)}
F_1\left(\frac{d-3}{2},1,\frac12; \frac{d-1}{2};\frac{tz}{4},1-\frac{t}{4m^2}
\right)
\label{solution}
\end{eqnarray}
with $z=\frac{4 u}{t(4 m^2-s)}$ and $s,t,u$ the usual Mandelstam variables.
Here the generalized hypergeometric functions are
\begin{eqnarray}
F_1\left(\frac{d-3}{2},1,\frac{1}{2},\frac{d-1}{2};x,y\right)=
&&      \sum_{r=0}^{\infty} \sum_{s=0}^{\infty} 
\frac{\left( \frac{d-3}{2} \right)_{r+s}}
{\left( \frac{d-1}{2} \right)_{r+s}}~\frac{\left( \frac{1}{2} \right)_s}
{\left( 1 \right)_s} ~x^r y^s,
\end{eqnarray}
\begin{eqnarray}
F_2\left(\frac{d-3}{2},1,1,\frac{3}{2},\frac{d-2}{2};x,y\right)=
&&      \sum_{r=0}^{\infty} \sum_{s=0}^{\infty} 
\frac{\left( \frac{d-3}{2} \right)_{r+s}}
{\left( \frac{3}{2} \right)_r \left( \frac{d-2}{2} \right)_s} ~x^r y^s 
\end{eqnarray}
and the Kamp\'e de F\'eriet function \cite{AppellKdF}
\begin{eqnarray}
F^{1;2;1}_{1;1;0} \left[^{\frac{d-3}{2}:~ \frac{d-3}{2},~1;~~~~ 1;}
_{\frac{d-1}{2}:~~~~~~ \frac{d-2}{2};~~-;}~~x,y\right]=
&&      
\sum_{r=0}^{\infty} \sum_{s=0}^{\infty} 
\frac{\left( \frac{d-3}{2} \right)_{r+s}}
{\left( \frac{d-1}{2} \right)_{r+s}}~\frac{\left( \frac{d-3}{2} \right)_r}
{\left( \frac{d-2}{2} \right)_r} ~x^r y^s = \phi(x,y).
\end{eqnarray}

$b_4$ is the boundary term, which we have to determine in the
following. First of all we calculate the derivatives $\partial_k \Delta_4$:
\begin{equation}
\partial_1 \Delta_4=\partial_3 \Delta_4=2st(4m^2-s),
\end{equation}
and 
\begin{equation}
\partial_2 \Delta_4=\partial_4 \Delta_4=-2 s t^2.
\end{equation}
These two sets of derivatives obviously have opposite sign
for physical values of the kinematical variables $s$ and $t$
(and also under their exchange for the crossed diagram).
Thus, according to the discussion after (\ref{xbarli}), we conclude that 
the boundary term is zero.

Nevertheless it is of interest to investigate
an alternative approach to determine the boundary term:
instead of the asymptotic method described in Sect. 3
we set up a differential equation for $b_4$ with respect to $m^2$ .
An expression for the derivative
of $I_{1111}^{(d)}$ with respect to $m^2$ can be obtained by
differentiating the parametric representation of $I_{1111}^{(d)}$
\begin{equation}
I_{1111}^{(d)}=\frac{1}{i^{d/2}}
\int_0^{\infty} ... \int_0^{\infty}
\prod_{j=1}^4  \frac{ d\alpha_j}{D^{\frac{d}{2}}}
~~\exp \left[ i\frac{Q}{D}-i\sum_{k=1}^4\alpha_k m_k^2 \right]
\label{intrep}
\end{equation}
where
\begin{eqnarray}
&&Q=\alpha_1\alpha_2 p_{12}^2
   +\alpha_1\alpha_3 p_{13}^2
   +\alpha_1\alpha_4 p_1^2
   +\alpha_2\alpha_3 p_{23}^2
   +\alpha_2\alpha_4 p_2^2
   +\alpha_3\alpha_4 p_3^2,
\nonumber \\
&&D=\alpha_1+\alpha_2+\alpha_3+\alpha_4.
\end{eqnarray}
Substituting the kinematical variables (\ref{Bhabkiva})
gives
\begin{equation}
Q- D ~\sum_{k=1}^4\alpha_k m_k  = \alpha_1 \alpha_3 t+ \alpha_2\alpha_4 s
-(\alpha_2+\alpha_4)^2 m^2.
\end{equation}
Differentiating (\ref{intrep}) we obtain
\begin{equation}
\frac{\partial I_{1111}^{(d)}}{\partial m^2}
=-2 I_{1311}^{(d+2)}-2 I_{1212}^{(d+2)}-2 I_{1113}^{(d+2)}.
\end{equation}
Using recursions with respect to the indices and dimension
as described in \cite{fjt}, the integrals on the right hand side can 
be reduced to master integrals with the result
\begin{eqnarray}
&&
\frac{\partial I_{1111}^{(d)}}{\partial m^2}
   =\frac{1}{ u (4m^2-s)}\left\{
       2 [(4-d)t -u]   I_{1111}^{(d)}
         - 8 (d-3)   I_F^{(d)} 
\right.
\nonumber \\
&&\left.
~~~~~~~~~~+\frac{2(d-2)}{m^2}I_1^{(d)}(m^2)
-\frac{2 u(d-3)}{t m^2 } I_G^{(d)} -(d-4) 
\frac{(u+4m^2)}{ m^2 }   J_G^{(d)}
\right \},
\label{i4fromQ}
\end{eqnarray}
where $I_F^{(d)}$ and $I_G^{(d)}$ are 2-point functions
\begin{equation}
I_G^{(d)}= \left. I_2^{(d)} \right|_{m_1=0,m_2=0}
=\frac{ \sqrt{\pi} }{(-p^2)^{(2-\frac{d}{2})}}~
\frac{\Gamma \left(2-\frac{d}{2} \right)
      \Gamma \left(\frac{d}{2}-1 \right)}
      {2^{d-3} ~\Gamma \left(\frac{d-1}{2}\right)},
\end{equation}
\begin{equation}
I_F^{(d)}=\left. I_2^{(d)}\right|_{m_1=m,m_2=m}
 =(m^2)^{(\frac{d}{2}-2)}~
 \Gamma \left( 2- \frac{d}{2} \right) \Fh21\FfP{1,2-\frac{d}{2}}{\frac32}.
\end{equation}

Differentiating (\ref{solution})
w.r.t. $m^2$ and equating the two expressions for the derivative 
$\frac{\partial I_{1111}^{(d)}}{\partial m^2}$, substituting 
(\ref{solution}) into the r.h.s. of  (\ref{i4fromQ}) yields a 
differential equation for $b_4$. To derive this equation we used 
\begin{eqnarray}
F_1&=&F_1\left(\frac{d-2}{2},1,\frac12,\frac{d}{2};x,y\right),
\nonumber \\
F_2&=&F_2\left(\frac{d-3}{2},1,1,\frac{3}{2},\frac{d-2}{2};x,y\right),
\nonumber \\
2(x-y) x \frac{d F_1}{dx} &=&
[x(2-d)+y(d-3)] F_1 
-y(d-3)F_1\left(\frac{d-2}{2},0,\frac12,\frac{d}{2};x,y \right)
\nonumber \\
&+&~~
x(1-y)(d-2)F_1\left(\frac{d}{2},1,\frac12,\frac{d}{2};x,y \right),
\nonumber \\
2(x-y) x \frac{d F_1}{dy} &=& yF_1 +y(d-3)
F_1\left(\frac{d-2}{2},0,\frac12,\frac{d}{2};x,y \right)
\nonumber \\
&-&~~y(1-x)(d-2)F_1\left(\frac{d}{2},1,\frac12,\frac{d}{2};x,y\right),
\nonumber \\
F_1(\alpha,0,\beta',\gamma;x,y)&=&~_2 F_1(\alpha,\beta',\gamma,y),
\nonumber \\
F_1(\alpha,\beta,\beta',\alpha;x,y)&=&\frac{1}
{(1-x)^{\beta}(1-y)^{\beta'}}.
\end{eqnarray} 

For the Kamp\'e de F\'eriet series the following relation was used
\begin{eqnarray}
&&2(y-x)x\frac{\partial \phi(x,y)}{\partial x}
= [y -(d-3)(y-x)] \phi(x,y)
\nonumber \\
&&-x(d-3)_2F_1\left(1,\frac{d-3}{2}, \frac{d-2}{2},x \right)
+(d-4)y~_2F_1\left(1,\frac{d-3}{2}, \frac{d-1}{2},y \right).
\end{eqnarray}
For the $F_2$ function we used the relation:

\begin{eqnarray}
&&  -8 m^2 s t u
       \frac{\partial F_2}{\partial x}  =
-8m^2(4m^2(t+u)-us )u
       \frac{\partial F_2}{\partial y}  
\nonumber \\       
&&~~~ + t(4 m^2-s) 
[( 4m^2(t+u)-u s ) (d-4)  - 4 m^2 u ] F_2
\nonumber \\
&&~~~+tu (s-8 m^2)
(s-4m^2)~_2F_1 \left(1,\frac{d-3}{2}, \frac{d-2}{2},- m^2 z \right)
\nonumber \\
&&~~~-(d-4)~t^2
(s-4m^2)^2~_2F_1\left(1,3-\frac{d}{2},\frac32, \frac{s}{4m^2}\right)
\end{eqnarray}

Finally the following differential equation was obtained for $b_4$:
\begin{equation}
\frac{\partial b_4}{\partial m^2} =\frac{2(u-4t)}
{u (s-4m^2)} b_4 .
\end{equation}
As initial value we chose $b_4(m^2=0)=0$, as we found explicitely
and thus $b_4=0$ for also for $m^2 \neq 0$.

%%%%%%%%%%%%%%%%%%%%%%%%%%%%%%%%%%%%%%%%%%%%%%%%%%%%%%%%%
%%%%%%%%%%%%%%%%%%%%%%%%%%%%%%%%%%%%%%%%%%%%%%%%%%%%%%%%%
\section{Conclusion}
%%%%%%%%%%%%%%%%%%%%%%%%%%%%%%%%%%%%%%%%%%%%%%%%%%%%%%%%%
%%%%%%%%%%%%%%%%%%%%%%%%%%%%%%%%%%%%%%%%%%%%%%%%%%%%%%%%%

   For many two-loop problems in the electroweak theory it
is necessary to calculate one-loop diagrams in arbitrary
dimension, e.g. to obtain the $\varepsilon$ expansion when 
inserting counter- terms. For this problem we offer a general
solution for $2-$, $3-$ and $4-$point functions for arbitrary
masses and kinematics. The solution is obtained in terms
of generalized hyper- geometric functions plus a `boundary
term'. To determine the latter, two different methods have
been applied: an asymptotic expansion in the dimension $d$
and alternatively a first order differential equation.
Our results (for arbitrary kinematics) seem fairly lengthy, 
but they are finally 
expressed in terms of functions $_2F_1$,$F_1$ and $F_S$,
which are quite accessible analytically as well as numerically.
In particular we point out the simple integral 
representations (\ref{intrepF1}) and (\ref{intrepFS}),
which can easily be evaluated for arbitrary kinematics, and thus
will become of particular importance for the evaluation of
$5-$point functions - as mentioned in the introduction.
For diagrams occurring in Bhabha scattering we have given
explicit results for arbitrary dimension of the needed
one-loop diagrams. The obtained results will serve in further
calculations as useful tools. 

%%%%%%%%%%%%%%%%%%%%%%%%%%%%%%%%%%%%%%%%%%%%%%%%%%%%%%%%%
%%%%%%%%%%%%%%%%%%%%%%%%%%%%%%%%%%%%%%%%%%%%%%%%%%%%%%%%%
\section{Acknowledgment}
%%%%%%%%%%%%%%%%%%%%%%%%%%%%%%%%%%%%%%%%%%%%%%%%%%%%%%%%%
%%%%%%%%%%%%%%%%%%%%%%%%%%%%%%%%%%%%%%%%%%%%%%%%%%%%%%%%%

We are thankful to J.~Gluza, M.~Kalmykov and T.~Riemann for useful 
discussions and remarks.
J.F. and O.V.T. are greatful to DESY for repeated invitations.
J.F. also acknow- ledges support by the European Community's 
Human Potential Programme under contract HPRN-CT-2000-00149,
Physics at Colliders. The authors also acknowledge support by the
DFG via the SFB/TR9-03.

\end{document}